\documentclass{emulateapj}
\usepackage{lscape}

\slugcomment{Draft}

\shorttitle{Spatially Resolved Star Formation Law in NGC 5194}
\shortauthors{Blanc et al.}

\begin{document}

\title{The Spatially Resolved Star Formation Law from Integral
  Field Spectroscopy:    \ \ \ \ VIRUS-P Observations of NGC 5194}

\author{Guillermo A. Blanc,\altaffilmark{1} Amanda
  Heiderman,\altaffilmark{1} Karl Gebhardt,\altaffilmark{1} Neal
  J. Evans II,\altaffilmark{1} Joshua Adams\altaffilmark{1}}
\altaffiltext{1}{Astronomy Department, University of Texas at Austin, USA}

\begin{abstract}

We investigate the relation between the star formation rate
surface density ($\Sigma_{SFR}$) and the mass surface density of gas
($\Sigma_{gas}$) in NGC 5194 (a.k.a. M51a, Whirlpool Galaxy). VIRUS-P
integral field spectroscopy of 
the central 4.1 $\times$ 4.1 kpc$^2$  of the galaxy is used to measure
H$\alpha$, H$\beta$, [NII]$\lambda\lambda$6548,6584, and [SII]$\lambda
\lambda$6717,6731 emission line fluxes for 735 regions $\sim$170 pc in
diameter. We use the Balmer decrement to calculate nebular dust
extinctions, and correct the observed fluxes in
order to measure accurately $\Sigma_{SFR}$ in each region. Archival
HI 21cm and CO maps with similar spatial resolution to that
of VIRUS-P are used to measure the atomic and molecular gas surface
density for each region. We present a new method for fitting the Star
Formation Law (SFL), which includes the intrinsic scatter in the
relation as a free parameter, allows the inclusion of non-detections in
both $\Sigma_{gas}$ and $\Sigma_{SFR}$, and is free of the systematics
involved in performing linear correlations over incomplete data in
logarithmic space. After rejecting regions whose nebular spectrum is
affected by the central AGN in NGC 5194, we use the [SII]/H$\alpha$ 
ratio to separate spectroscopically the contribution from the diffuse
ionized gas 
(DIG) in the galaxy, which has a different
temperature and ionization state from those of H II regions in the
disk. The DIG only accounts for 11\% of the
total H$\alpha$ luminosity integrated over the whole central region,
but on local scales it can account for up to a 100\% of the H$\alpha$
emission, especially in the inter-arm regions. After removing the DIG
contribution from the H$\alpha$ fluxes, we measure a slope
$N=0.82\pm0.05$, and an intrinsic scatter $\epsilon=0.43\pm0.02$ dex
for the molecular gas SFL. We also measure a typical depletion timescale $\tau
=\Sigma_{HI+H_2}/\Sigma_{SFR} \approx 2$ Gyr, in good agreement with
recent measurements by \cite{bigiel08}. The atomic gas density
shows no correlation with the SFR, and the total gas SFL in the
sampled density range closely follows the molecular gas SFL.
Integral field spectroscopy allows a much cleaner
measurement of H$\alpha$ emission line fluxes than narrow-band
imaging, since it is free of the systematics introduced by continuum
subtraction, underlying photospheric absorption, and
contamination by the [NII] doublet. We assess the validity of
different corrections usually applied in narrow-band measurements to
overcome these issues and find that while systematics are introduced
by these corrections, they are only dominant in the low
surface brightness regime. 
The disagreement with the previous measurement of a
super-linear molecular SFL by \cite{kennicutt07} is most likely due to
differences in the 
fitting method. Our results support the recent evidence for a low, 
and close to constant, star formation efficiency (SFE=$\tau^{-1}$) in
the molecular component of the ISM. 
The data shows an excellent agreement
with the recently proposed model of the SFL by \cite{krumholz09b}.
The large intrinsic scatter observed may imply the existence of other
parameters, beyond the availability of gas, which are important at
setting the SFR. 

\end{abstract}

\keywords{galaxies:}

\section{Introduction}

In the quest to achieve a thorough understanding of the processes
involved in the formation and subsequent evolution of galaxies, we
must first fully characterize the process of star formation under different
environments in the ISM. During the last decade, major efforts have been
made to characterize the variables involved in triggering star
formation and setting the star formation rate (SFR) in
galaxies. \cite{kennicutt98b} showed that, integrating over the whole 
optical disk of galaxies, the star formation
rate surface density ($\Sigma_{SFR}$), as measured by the H$\alpha$
emission, tightly correlates with the total gas surface density 
($\Sigma_{HI+H_2}$) over several orders of magnitude
in SFR and gas density. The relation from Kennicutt follows a
power-law form, with a slope 
$N=1.4$. These types of correlations between $\Sigma_{SFR}$ and
$\Sigma_{gas}$, either 
atomic ($\Sigma_{HI}$), molecular ($\Sigma_{H_2}$), or total
($\Sigma_{HI+H_2}$), are usually known as Star
Formation Laws  \citep[SFL, a.k.a. Schmidt Laws or Schmidt-Kennicutt Laws,
after][who first introduced the power-law
parametrization to relate gas density and the SFR]{schmidt59}, and
they show that 
the availability of gas is a key variable in setting the SFR.

Although the global SFL provides us with valuable insights on the
role that gas density plays at setting the SFR, the measurement
involves averaging over the many orders of magnitude in
$\Sigma_{gas}$ and $\Sigma_{SFR}$ present in the ISM of single
galaxies, implying the loss of valuable information about the detailed
physics that give rise to the SFL. Azimuthally averaged measurements
of gas surface densities and the SFR have been used to conduct more detailed
studies of the SFL across the disks of local galaxies. For example,
\cite{wong02} measured, under the assumption of constant dust
extinction, a slope of $N\approx 0.8$ for the
molecular SFL, and $N \approx 1.1$ for the total gas SFL
on a sample of seven molecule rich spirals, with a large scatter from
galaxy to galaxy, and \cite{schuster07} measured
$N=1.4\pm0.6$ for the total gas SFL on NGC 5194. Azimuthally averaged
profiles are 
also affected by averaging effects since $\Sigma_{SFR}$ and
$\Sigma_{gas}$ can change by more than 2 orders of magnitude at
constant galactocentric radius due to the presence of spiral structure.
We refer the reader to \cite{bigiel08} for a thorough compilation 
of previous measurements of the SFL in local galaxies.

More recently two studies have been aimed at measuring the ``spatially
resolved'' SFL
throughout the disks of nearby galaxies. \cite{kennicutt07} used a
combination of narrow-band H$\alpha$ and 24$\mu$m photometry to
estimate $\Sigma_{SFR}$, as well as 21cm and CO J=1-0 maps to
measure $\Sigma_{gas}$ for 257 star-forming regions, 520 pc in
diameter, in the disk of NGC 5194. They measured slopes of
$N=1.37\pm0.03$ and $N=1.56\pm0.04$ for the molecular
and total gas SFL respectively. \cite{bigiel08} used far-UV and
24$\mu$m images to create a $\Sigma_{SFR}$ map, and 21cm, CO
J=2-1, and CO J=1-0 data to create $\Sigma_{gas}$ maps of seven
spiral galaxies and eleven late-type/dwarf galaxies. After convolving
the maps to a common resolution of 750 pc, they performed a
pixel-to-pixel analysis and measured a molecular SFL with an average $N=1.0\pm0.2$
for the normal spirals ($N=0.84$ for NGC 5194). Both studies found a
lack of correlation 
between the SFR and the atomic gas density, which saturates around a
value of $10{\rm \; M_{\odot}pc^{-2}}$. This value is thought to be
associated with a density threshold for the formation of molecular
gas, and is consistent with predictions from theoretical modeling of
giant atomic-molecular complexes \citep{krumholz09a}. The
total gas SFL is then driven by the correlation between the molecular
gas density and the SFR, and the molecular fraction in the ISM. At
the highest densities present in normal spiral galaxies 
($\Sigma_{HI+H_2}=50-1000{\rm\; M_{\odot}pc^{-2}}$) the ISM is mostly
molecular and the total gas SFL closely follows the H$_2$ SFL. At
densities lower than $10{\rm M_{\odot}pc^{-2}}$ the total gas SFL gets
much steeper due to a strong decrease of the molecular fraction. This
behavior has been recently modeled by \cite{krumholz09b}.

While spatially resolved studies of the SFL obtain consistent
results on the behavior of the atomic gas, they disagree when it
comes down to the molecular component. The \cite{bigiel08} measurement 
of a linear molecular SFL is consistent with a scenario in which 
star formation occurs at a constant efficiency inside GMCs, whose
properties are fairly uniform across normal spiral galaxies
\citep{blitz07, bolatto08}. This
homogeneity in the properties of GMCs is expected if they are
internally regulated by processes like stellar feed-back, and they are
decoupled from their surroundings due to the fact of being
strongly overpressured \citep{krumholz09b}. \cite{kennicutt07}, on the other
hand, measured a super-linear molecular SFL in NGC 5194, which
suggests an increasing SFE towards higher gas densities. Although
the authors state that a super-linear slope ($N>1$) is still consistent with a
constant ``efficiency'' if the star-forming lifetimes of massive clouds
were systematically lower than those of low-mass clouds, this is true
only if the efficiency is defined as the ratio of the produced stellar
mass over the available molecular gas mass, which is the classical
definition used by galactic studies in the Milky Way. In this work, as well as
in \cite{bigiel08}, the efficiency is defined as
SFE$=\Sigma_{SFR}/\Sigma_{gas}$, or the inverse of the depletion time, 
so shorter star formation timescales imply a higher SFE, and a
super-linear SFL always translate in higher SFE at higher gas densities. 

With the goal of investigating this issue, we have
conducted the first measurement of the spatially resolved SFL
using integral field spectroscopy. We mapped the H$\alpha$
emission in the central 4.1$\times$4.1 kpc$^2$ of the nearby face-on
spiral galaxy NGC 5194 using the
Visible
Integral field Replicable Unit Spectrograph Prototype 
\citep[VIRUS-P,][]{hill08}. Hydrogen
recombination lines are known to be good tracers of the SFR. Their
intensity scales linearly with the ionizing UV flux in galaxies, which
is dominated
by the emission from massive stars ($\geq 10 \rm{\; M_{\odot}}$) with typical
lifetimes of $< 20$ Myr, hence they provide an almost instantaneous
measurement of the SFR \cite[and references therein]{kennicutt98a}. 

Due to the small field of view of current integral field units (IFUs),
typically less than 1 arcmin$^2$, 2D spectroscopic H$\alpha$ mapping of 
nearby galaxies with large angular sizes has not been conducted
efficiently in the 
past. Instead, narrow-band imaging has been typically used to
construct H$\alpha$ based SFR maps. H$\alpha$ narrow-band imaging suffers
from contamination from the [NII]$\lambda\lambda$6548,6584 doublet,
and is sensitive to systematic errors in continuum subtraction
and the estimation of the strength of the H$\alpha$ absorption in the
underlying stellar spectrum. Spectroscopic measurements are free of
all these sources of error, and hence provide a much cleaner
measurement of H$\alpha$ fluxes. A major part of this paper is
dedicated to investigate these systematics in order to assess the
validity of the typical corrections applied to narrow-band H$\alpha$ images.

VIRUS-P is the largest
field of view IFU in the world and it allows for efficient H$\alpha$ mapping
of nearby galaxies. The observations presented here were taken as part of
the VIRUS-P Exploration of Nearby Galaxies
(VENGA\footnote{http://www.as.utexas.edu/$\sim$gblancm/venga.html},
Blanc et al. in 
preparation). VENGA is a large scale extragalactic IFU survey that
will spectroscopically map large parts of the disks of $\sim 20$
nearby spirals, to allow a number of studies on
star-formation, structure assembly, stellar populations, gas and
stellar dynamics, chemical evolution, ISM structure, and galactic
feedback.

\begin{figure*}[t]
\begin{center}
\plotone{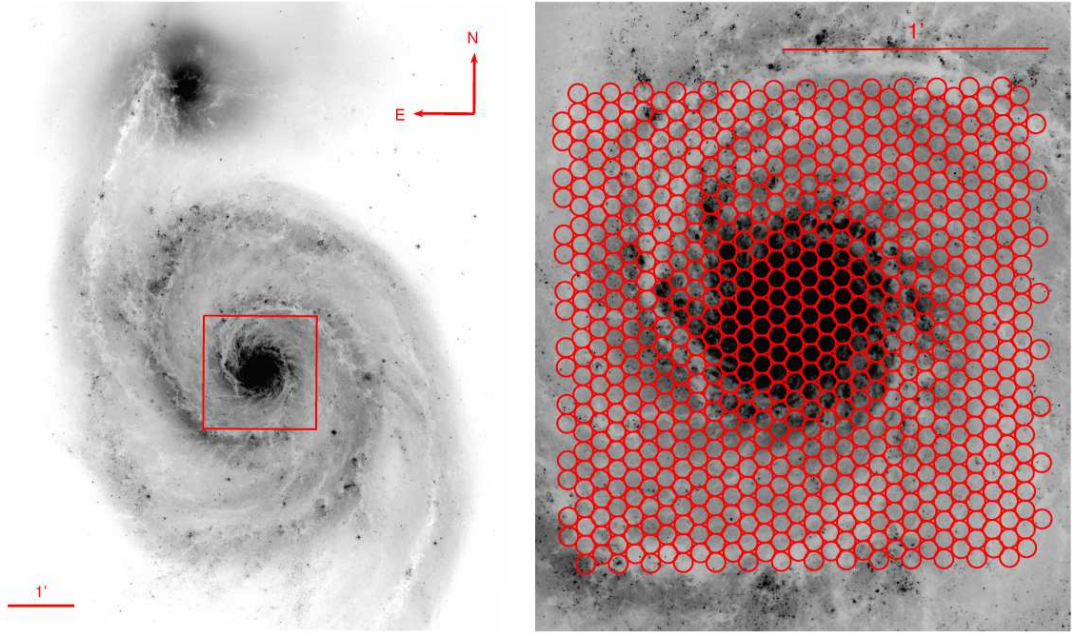}
\caption{{\it Left:} HST+ACS V-band image of NGC5194 and its companion
  NGC 5195 \citep{mutchler05}. The central $4.1 \times 4.1$ kpc$^2$ region sampled by the
  $1.7' \times 1.7'$ VIRUS-P field of view is marked in
  red. {\it Right:} Map of the 738 regions sampled by VIRUS-P in the 3
  dither positions. Each region has a diameter of 4.3$''$
  corresponding to $\sim$170 pc at the distance of NGC5194.}
\label{fig-1}
\end{center}
\end{figure*}

The VIRUS-P spectral map was used in combination with CO J=1-0 and
HI 21cm intensity maps of NGC 5194 from the BIMA Survey of Nearby
Galaxies, SONG 
\citep{helfer03}, and The HI Nearby Galaxy Survey, THINGS
\citep{walter08}, to measure $\Sigma_{SFR}$,
$\Sigma_{H_2}$, and $\Sigma_{HI}$ in order to construct the spatially 
resolved SFL. In \S 2 and \S 3 we present the VIRUS-P observations and 
the data reduction and calibration methods. In \S 4 we describe the CO
and 21cm data used to measure the molecular and atomic gas surface densities,
as well as a HST NICMOS Pa$\alpha$ image used to validate our
dust extinction measurements. \S 5 presents the methods used
to remove the stellar continuum and measure accurate nebular emission
line fluxes, together with our dust extinction correction. 
The calculation of $\Sigma_{gas}$ is described in \S 6. The rejection
of regions whose nebular emission is affected by the central AGN in
NGC 5194 is presented in \S 7. The correction to account for the
contribution of the DIG to the H$\alpha$ fluxes is described in
\S8. The resulting 
spatially resolved SFLs for the molecular, atomic and total gas are
presented in \S 9, followed by a 
discussion on the implications of our results for narrow-band imaging
surveys in \S 10. Finally we compare our results with previous
measurements and
theoretical predictions of the SFL in \S 11, and present our
conclusions in \S 12.

Throughout this paper we assume a distance to NGC 5194 of 8.2 Mpc for
consistency with \cite{kennicutt07}. While \cite{bigiel08} used a
slightly smaller distance of 8.0 Mpc, it is worth noticing that most
of the results in this paper are based on surface densities, which are
independent of distance, and thus are not affected by the assumed
value. All values for $\Sigma_{SFR}$ are in units of ${\rm M_{\odot}
  yr^{-1 } kpc^{-2}}$, and values of
$\Sigma_{gas}$ are in units of ${\rm M_{\odot}pc^{-2}}$.

\section{Observations}

We obtained spatially resolved spectroscopy over the central
4.1$\times$4.1 kpc$^2$ 
region of NGC 5194 on the night of April 4, 2008, using VIRUS-P
on the 2.7m Harlan J. Smith telescope at McDonald
Observatory. VIRUS-P with the VP-2 IFU bundle used in this work consists of a
square array of 246 optical fibers which samples a $1.7'\times 1.7'$
field of view with a 1/3 filling factor. The fibers are 200$\mu$m in
diameter, corresponding to 4.3$''$ on sky. The spectrograph images the
spectrum of the 246 fibers on a 2048$\times$2048 Fairchild Imaging
CCD. Because of camera
alignment issues, the spectrum of one fiber fell off the chip, reducing
the number of usable fibers to 245.

The spectrograph was used in a red setup under which it samples a
wavelength range of 4570-6820\AA\ with a spectral resolution of
$\sim$5.0\AA\ (FWHM). This red setup allows us to sample both H$\beta$
and H$\alpha$, and our resolution is high enough to resolve the
H$\alpha$-[NII]$\lambda\lambda$6548,6893 complex. We took the data in
2x1 binning mode in the spectral direction which translates into a
plate scale of 2.2 \AA pixel$^{-1}$. Given the 1/3 filling factor of
the IFU, three dithered exposures were necessary to sample the complete
field of view.

We obtained four 20 minute exposures at each of the 3 dither
positions, accounting for an effective exposure time of 80 minutes. 
Dither 1 was centered at
$\alpha$=13:29:52.69;$\;\delta$=+47:11:43.0. Dithers 2 and 3 were
offset from dither 1 by $\Delta \alpha=-3.6''$;$\;\Delta
\delta=-2.0''$ and $\Delta \alpha=0.0''$;$\;\Delta \delta=-4.0''$
respectively. Figure \ref{fig-1} shows the observed region in NGC 5194
as well as the position of the IFU fibers for the 3 dithers. Because
of the extended 
nature of NGC 5194 no fibers in the field of view sampled a blank
region of the sky. This implied the need for off-source sky frames in
between science frames. We obtained 5 minute sky exposures bracketing
all science exposures. These were obtained 30$'$ North of NGC
5194. The typical seeing during the observations was 2.0''.

Bias frames, comparison NeCd lamps, and twilight flats were taken at
the beginning and end of the night. VIRUS-P is mounted on a two-degree
of freedom gimble at the broken cassegrain focus of the telescope. The
gimble keeps the spectrograph in a fixed gravity vector independent of
the position of the telescope during the observations which translates
into a practically complete lack of flexure in the spectrograph
optical components. For this reason calibration frames intercalated
with the science observations were not necessary.

\begin{figure*}[ht]
\begin{center}
\plotone{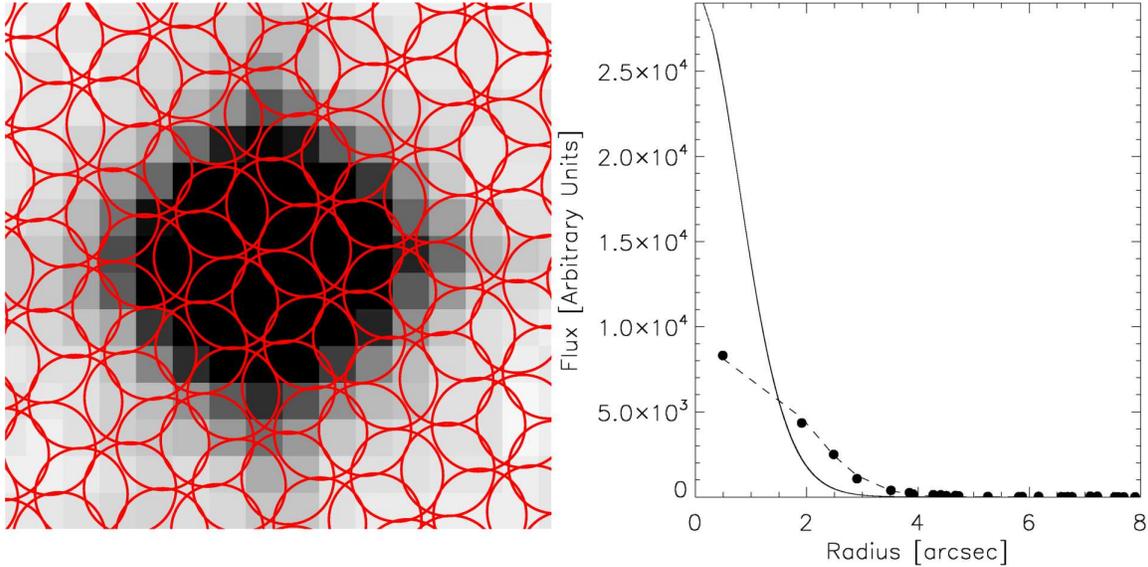}
\caption{{\it Left:} DSS image of Feige 34. Superimpossed is the 6
  dither position pattern used to observe spectro-photometric
  standard stars. {\it Right:} Flux measured by each fiber as a
  function of its distance to the PSF centroid (filled circles). Also
  shown are the best-fitted Moffat PSF (solid line), and its
  fiber-sampled light distribution (dashed line).}
\label{fig-2}
\end{center}
\end{figure*}

The spectro-photometric standard Feige 34 was observed for the purpose
of flux calibration (see \S 3.1). Standard observations were
performed using a finer 6 position dither pattern which better samples
the PSF of the star and ensures the collection of its total flux (see
\S 3.1 and Figure \ref{fig-2}).

The instrument is equipped with a guiding camera which images a $4.5'
\times 4.5'$ field offset from the science field sampled by the
IFU. The guiding camera is a $512\times 512$ pixel Apogee unit
equipped with a BV filter which allows broad-band photometric
measurements of the stars in the field. During the night we saved a
guider frame every 30 seconds in order to reconstruct
changes in atmospheric transparency. The guider images are also used
to establish the IFU astrometry. The relative offset, rotation and
plate scales of the guider and IFU fields have been precisely
calibrated using observations of crowded fields in open clusters, so
the pixel coordinates of stars in the guider frames provide us with
coordinates for the center of all fibers in the IFU with an
astrometric rms of $\sim$0.5$''$.

In this way we obtained spectra for 735 regions 4.3$''$ 
in diameter ($\sim$170 pc at the distance of NGC 5194), in the central
region of the galaxy. The spectra reaches a median 5$\sigma$ sensitivity
in continuum flux density of $2.5\times 10^{-17}$ erg
s$^{-1}$cm$^{-2}$\AA$^{-1}$, which translates into a median signal-to-noise
(S/N) ratio per resolution element of 95 (53 for the faintest fiber).

\section{Data Reduction}

Data reduction is performed using our custom pipeline
VACCINE (Adams et al. in preparation). Individual frames are overscan 
and bias subtracted, and bad
pixels are masked. We use the twilight flats to trace the peak of
the spatial profile of the spectrum of all fibers on the chip, and
extract the 2D spectrum of each fiber on the science frames,
comparison lamp frames,  
and flats using a seven pixel aperture around the peak.

The extracted comparison lamp spectra are used to compute an
independent wavelength solution for each fiber. We use 4th order
polynomials to compute the wavelength solutions which show a typical
rms of 0.2 \AA\ ($\sim$0.1 pixel).

We correct the twilight flats for solar absorption lines and use them to
measure the shape and amplitude of the spatial profile of the fibers
as a function of wavelength. This profile is given by the point spread
function (PSF) of the fibers on chip in the spatial direction, and the relative
instrumental throughput of each fiber as a function of
wavelength. Dividing the twilight flats by this profile yields a
pixel-to-pixel flat. We divide all science, sky, and spectro-photometric
standard frames by both the fiber profile and the
pixel-to-pixel flats. This removes any fiber-to-fiber and pixel-to-pixel 
variations in sensitivity.

A background frame is created for each science exposure by averaging
the two bracketing 5 minute sky frames and scaling by the difference in
exposure time. We estimate the sky spectrum for each fiber by fitting a
non-uniform spline to the spectra of the 60 neighboring
fibers on chip in the background frame. This spectrum is subtracted from each
fiber in the science data.

In order to test the quality of our background subtraction algorithm
we construct background frames for each of our sky exposures using
the two closest of the other sky exposures. We then follow the same
procedure to background subtract our sky frames. We observe residuals
centered around zero in the background subtracted sky frames that are
less than 1\% of the galaxy continuum flux in the faintest fibers in
our science data. The only exception are the regions of the spectra at
the wavelength of the 4 brightest sky emission lines in our wavelength
range in which the residuals can be considerably larger due to the
fast time variability of these spectral features. These regions
showing poor background subtraction are masked in our science
data. At this stage we combine individual exposures using a
biweight \citep{beers90}.

Error maps including Poisson photon count uncertainties and
 read-noise are created for every fiber on each frame. We use these error
maps together with the fiber profile to calculate 
the weights used for collapsing the 2D spectrum into a 1D spectrum. The flux in
photo-electrons at each wavelength after collapsing is given by

\begin{equation}
f_{\lambda}=\frac{\sum_{i=1}^7 \left(\frac{p_i}{e_i} \right)^2 G f_{\lambda,i}}{\sum_{i=1}^7 \left(\frac{p_i}{e_i} \right)^2}
\end{equation}

where $p_i$ is the value of the fiber profile, $G$ is the gain, $f_i$
is the flux in
ADUs in the combined background subtracted spectrum, and $e_i$ is the
corresponding
error at each pixel as measured in the error map. This is equivalent
to weighting the pixels by
(S/N)$^2$. The sum is performed at every wavelength (column)
over the the 7 pixel aperture used for extraction. The final product
is a wavelength calibrated 1D spectrum of the area sampled by each
of the 245 fibers on each of the 3 dither positions on the galaxy.

\subsection{Flux Calibration}

Flux calibration of IFU data can be challenging but, if proper
care is taken, very accurate spectro-photometry can be achieved. This
is mostly because of the lack of a wavelength dependent slit loss
function. Atmospheric dispersion can change the position of a standard
star in the field of view as a function of wavelength, but as long as
the field is completely sampled by the fibers the total flux of the star
at all wavelengths is always collected. Also, photometry of stars in
the guider images taken during the observations allows us to
measure and correct for changes in atmospheric transparency during the
night.

During the observation of standard stars, fibers in the IFU only
sample a region of the star's PSF. Determining the fraction of the
total flux collected by each fiber is essential in order to compute a
proper instrumental sensitivity function by comparing each fiber
spectrum to the total intrinsic spectrum of the star. This requires
knowledge of the shape of the PSF as well as the distance from the PSF
centroid to the center of each fiber.

The spectro-photometric standard star Feige 34 was observed using a 6
position dither pattern shown in the left panel of Figure \ref{fig-2}. This
tight pattern provides a better sampling of the PSF and ensures we
collect the total flux of the star. We calculate the position of the 
centroid of
the PSF relative to the fibers by taking the weighted
average of the fibers positions in the field of view, using the
measured flux in each of them as weights. This corresponds to the first
moment of the observed light distribution.

The filled circles in the right panel of Figure \ref{fig-2} show the
flux measured in each fiber as a function of its radial distance to the
PSF center. This information can be used to reconstruct the shape of
the star PSF at the moment of the observations. In order to do this,
we assume a Moffat profile for the PSF and reconstruct its observed
light distribution by summing the flux in 4.3'' diameter circular
apertures at the corresponding radial distance of each fiber. The best-fitted
PSF and its fiber sampled light distribution are shown by the solid
and dashed curves in the right panel of Figure \ref{fig-2}. It can be
seen that the best-fitted model PSF, after being sampled by the fibers in our
dither pattern, matches the measured flux remarkably. This PSF model
allows us to know what fraction of the star total flux was measured
by each fiber during the observations.

We normalize the spectrum of each fiber by the fraction of the total
flux it sampled, and average this value for all fibers having a
significant ($> 5\sigma$) flux measurement in order to obtain the
star total instrumental spectrum. We correct the total spectrum by
atmospheric extinction and use the Feige 34 measurement of
\cite{oke90} to construct our instrumental sensitivity function.

Relative variations in atmospheric transparency during the night are
measured by performing aperture photometry on stars in the guider
images. Observing conditions were confirmed to be very stable, with
maximum variations in transparency of less than a 10\%. All spectra
in our science frames are corrected by this variations, atmospheric
extinction, and flux calibrated using the instrumental sensitivity
function.

It is important to notice that any difference in illumination or
throughput between fibers was taken out during the flat-fielding process,
so a common sensitivity function applies to all fibers. Our final
product is a wavelength and flux calibrated spectrum for the
735 regions.

In order to estimate the systematic uncertainty in our flux
calibration we have compared sensitivity functions computed using
different standard star observations taken as part of different
observing programs with VIRUS-P. Comparison of 10 standards taken
between September 2007 and June 2008 under different observing
conditions show that after correcting for relative changes in
atmospheric transparency (using photometry of stars in the guider
images) the computed sensitivity functions show an rms scatter of less
than 5\%.

\begin{figure*}[ht]
\begin{center}
\plotone{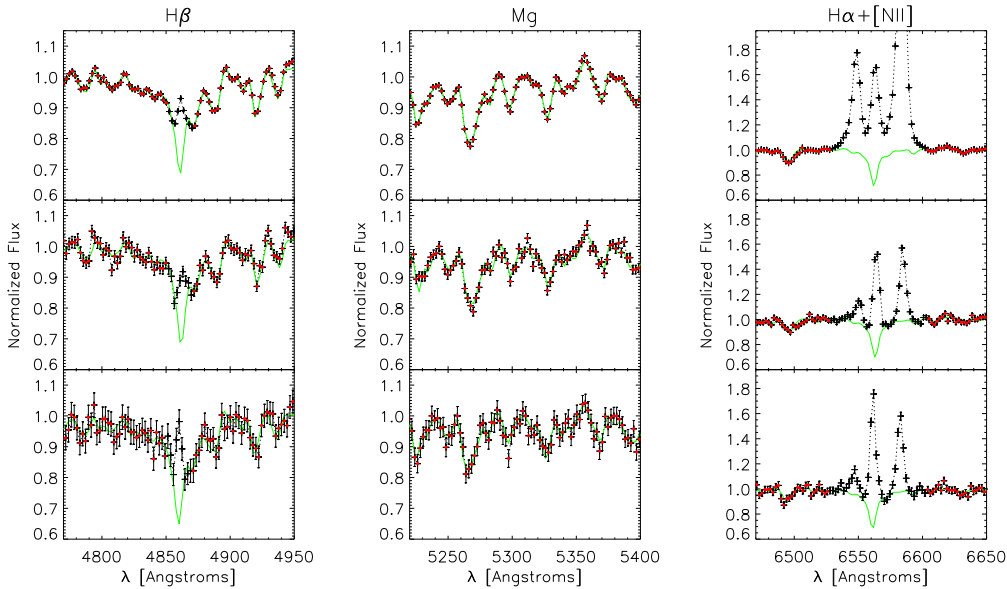}
\caption{Continuum normalized spectra around the H$\beta$, MgII, and
  H$\alpha$ features for 3 regions having the highest, median and
  lowest (top, middle, bottom) S/N per resolution element in the
  continuum. Crosses show the data with error bars. Red crosses mark
  the data points used to fit the best linear combination of stellar
  templates (green solid line). Black crosses were masked in the fit
  due to the presence of nebular emission.}
\label{fig-3}
\end{center}
\end{figure*}

\begin{figure*}[ht]
\begin{center}
\plotone{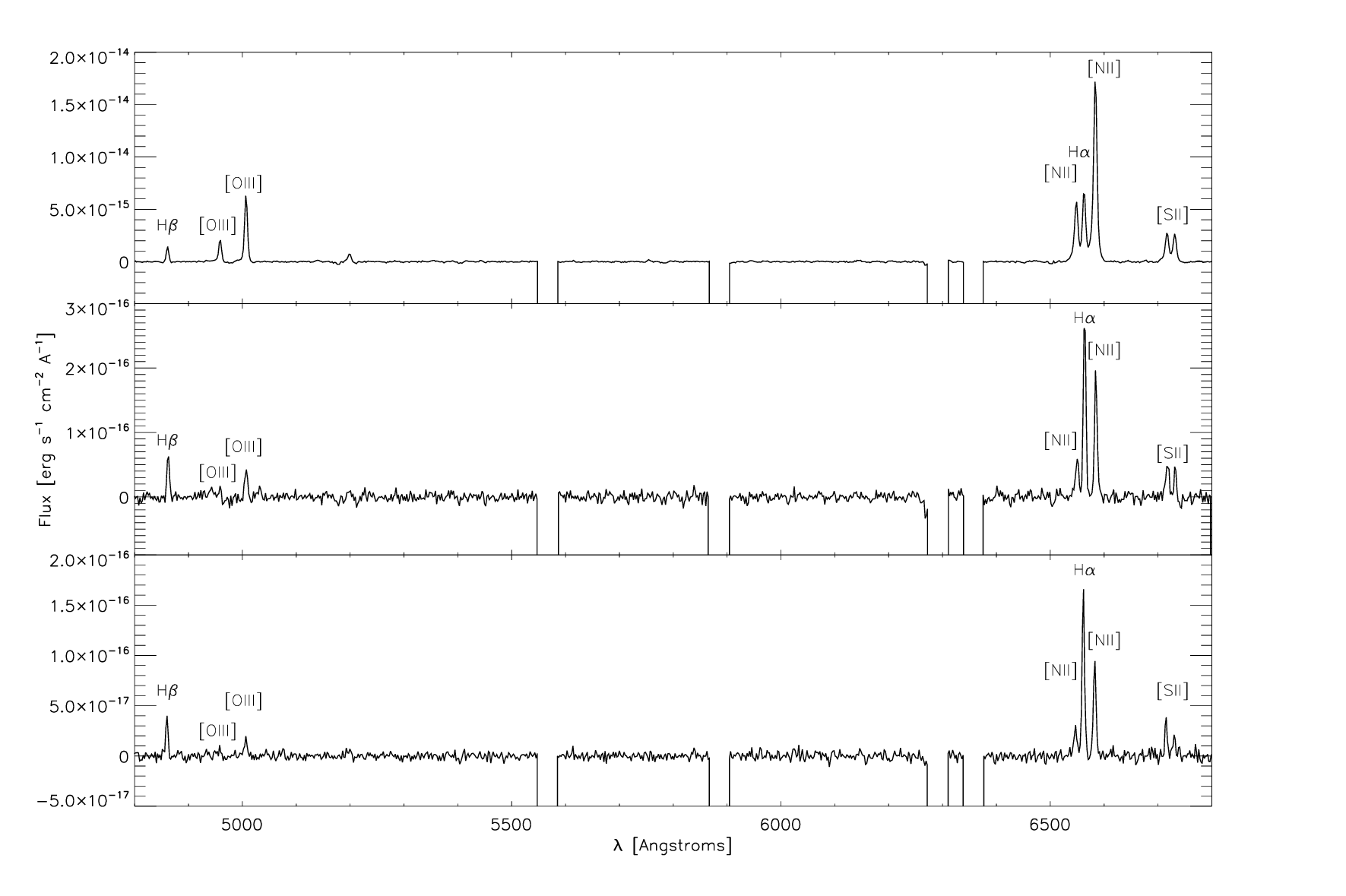}
\caption{Nebular emission spectrum of the same regions shown in Figure
\ref{fig-3}, obtained by subtracting the best-fitted linear combination
of stellar templates from the observed spectrum. Masked parts of the
spectra correspond to the regions around strong night sky emission
lines showing background subtraction residuals.}
\label{fig-4}
\end{center}
\end{figure*}

\section{Other Data}

\subsection{THINGS HI Data}

We use a combined 21 cm line intensity map of NGC 5194 from the Very Large
Array (VLA) taken as part of The HI Nearby Galaxy Survey
\citep[THINGS\footnote{http://www.mpia.de/THINGS/Data.html};][]{walter08}
 to estimate the atomic gas surface density
($\Sigma_{HI}$). HI data for NGC 5194 was taken using the B, C, and D
arrays during 2004 and 2005, with a combined on source integration time of
$\sim$10 hours. The final co-added (B+C+D array) integrated intensity
map has a robustly weighted beam size of $5.82''\times
5.56''$, which is well matched to the $4.3''$ VIRUS-P fiber diameter
convolved with the $2''$ seeing. The 1$\sigma$ noise per 5.2
km s$^{-1}$ channel is 0.44 mJy beam$^{-1}$ corresponding to a atomic
gas surface density of $\Sigma_{HI}$= 0.59 M$_{\odot}$ pc$^{-2}$. For
more details on data products and data reduction see \cite{walter08}.

\subsection{BIMA SONG CO Data}

Molecular gas surface densities are measured using the CO J=1-0 
intensity map of NGC 5194 from the Berkeley
Illinois Maryland Array (BIMA) Survey of Nearby Galaxies \citep[BIMA
SONG\footnote{http://nedwww.ipac.caltech.edu/level5/March02/SONG/SONG.html};
][]{helfer03}.  Zero spacing single dish data from the
NRAO 12 m telescope was combined with the interferometric
BIMA C and D array data, resulting in a map with a robust beam size of
$5.8''\times 5.1''$, well matched to the 21 cm map and the VIRUS-P
spatial resolution.  The corresponding 1$\sigma$ noise is 61 mJy beam$^{-1}$
in a 10 km s$^{-1}$ channel or $\Sigma_{H_2}$ = 13 M$_{\odot}$
pc$^{-2}$. For more details on the observations and the data reduction
refer to \cite{helfer03}.\\

\subsection{HST NICMOS Paschen-$\alpha$ Data}

The center of NGC 5194 was imaged in Pa$\alpha$ by \cite{scoville01}
using HST+NICMOS. A 3$\times$3 mosaic covering the central $186''
\times 188''$ of
the galaxy was imaged using the F187N and F190N narrow-band filters,
sampling the Pa$\alpha$ line and the neighboring stellar continuum
respectively. In this work we use this continuum subtracted Pa$\alpha$
image to
measure emission line fluxes to check the validity of our dust
extinction correction.  The data reduction, mosaicking, flux
calibration and continuum subtraction are described in
\cite{scoville01} and \cite{calzetti05}. The Pa$\alpha$ image overlaps
completely with the VIRUS-P pointing shown in Figure \ref{fig-1}.

\section{Measurement of Emission Line Fluxes}

We estimate the current $\Sigma_{SFR}$ for each region by means of the H$\alpha$
nebular emission luminosity. In this section we describe the methods
used to separate the emission lines coming from ionized gas from the
underlying stellar spectrum, measure emission line fluxes, and
estimate the dust extinction in each region using the
H$\alpha$/H$\beta$ ratio. 

\subsection{Photospheric Absorption Lines and Continuum Subtraction}

In galaxy spectra, both the H$\alpha$ and H$\beta$ emission lines sit
on top of strong Balmer absorption lines characteristic of the
photospheric stellar spectrum of young stars. Removing the
contribution from these absorption lines is essential in order to
estimate properly the emission line flux.

We use a linear combination of stellar template spectra to fit the
absorption line spectrum of each region. The templates are high S/N,
high resolution, continuum normalized spectra of a set of 18 stars
from the Indo-U.S. Library of Coud\'e Feed Stellar Spectra
\citep{valdes04}. Stars were chosen to span a wide range in spectral
types and metallicities (A7 to K0, and [Fe/H] from -1.9 to 1.6). 

The resolution of the templates is degraded
to match the VIRUS-P 5.0\AA\ spectral resolution. For each of the 735
regions, we mask the parts of the galaxy spectrum
affected by emission lines and sky subtraction residuals from bright
sky lines. The continuum at each wavelength is estimated using an
iterative running median filter, and used to normalized the observed
spectrum. 

We use this masked, continuum normalized spectrum to fit the best
linear combination of stellar templates for each region. Figure
\ref{fig-3} shows the best-fitted template combinations in regions centered in 
H$\beta$, Mg b, and H$\alpha$ for 3
regions in the galaxy. The bottom, middle and upper panels correspond
to fibers with the lowest, median and highest S/N in their
spectra respectively. For all 735 regions we obtain excellent fits to
the underlying
stellar spectrum. Figure \ref{fig-3} shows the importance of taking
into account the effect of photospheric Balmer absorption lines when
measuring H$\alpha$ and H$\beta$ fluxes. Ignoring the presence of
the absorption features can introduce serious underestimations of the 
emission line fluxes. For H$\alpha$ this effect can account for
underestimations of up to 100\% as will be shown in \S 10.

The best-fitted linear combination of stellar templates is scaled by the
galaxy continuum and subtracted from the original
spectrum in order to produce pure nebular emission line spectra for all
fibers. Figure \ref{fig-4} shows the nebular spectrum of the same
regions shown in Figure \ref{fig-3}. After subtracting the stellar
light, we are able to identify most well known emission features in
galaxy spectra. H$\beta$, [OIII]$\lambda \lambda $4959,5007,
[NII]$\lambda \lambda$,6548,6584, H$\alpha$ and [SII]$\lambda
\lambda$,6717,6731 are clearly seen in the spectra of all 735
regions. Visual inspection of Figure \ref{fig-4} shows that the
[NII]$\lambda \lambda$,6548,6584/H$\alpha$ ratio can change
drastically from region to region. This effect can introduce
systematic biases in narrow-band measured H$\alpha$ fluxes if the ratio is
assumed to be constant across the disk \citep{calzetti05,
kennicutt07}. This issue will be discussed in detail in \S 10.

\subsection{Emission Line Fluxes}

We measure emission line fluxes by independently fitting
H$\beta$, the H$\alpha$-[NII]$\lambda
\lambda$,6548,6584 complex, and the [SII]$\lambda \lambda$,6717,6731
doublet. Although the lines in the H$\alpha$-[NII] complex are clearly
resolved in our spectra, their wings show some level of overlap so we
used a 3 Gaussian component model to fit these lines. Similarly a 2
Gaussian component model was used to fit the [SII] doublet. H$\beta$
was fitted using a single Gaussian. These fits
provide the total flux and its uncertainty of all the above lines for the
735 regions. All lines are detected with a significance higher than
3$\sigma$ in all fibers. We measure a median and lowest S/N over all
fibers of 109 and 15 for H$\alpha$, 29 and 4 for H$\beta$, 49 and 13 for
[NII]$\lambda$6584, and 32 and 5 for [SII]$\lambda$6717. Emmision
line fluxes of all lines for all fibers are given in Table 1,
available in the electronic version of this paper.

\subsection{Extinction Correction from the Balmer Decrement}

The observed spectra is affected by differential extinction due to the
presence of
dust in the ISM of both NGC 5194 and the Milky Way. Before attempting
to estimate SFRs from H$\alpha$ fluxes, these have to be corrected for
dust extinction. Failing to do so can introduce underestimations in
the SFR of up to factors of $\sim$10 in the regions we are
studying. The Balmer line ratio H$\alpha$/H$\beta$, as will be shown
bellow, provides a good estimate of the dust extinction 
at the wavelength of the H$\alpha$ line.

Assuming an intrinsic H$\alpha$/H$\beta$ ratio of 2.87
\citep{osterbrock06}, the observed ratio provides the extinction at
the wavelength of H$\alpha$ through the following equation,

\begin{equation}
A_{H\alpha}=-2.5\;{\rm log}\left[ \frac{[H\alpha/H\beta]_{obs}}{2.87}\right]
\left(\frac{1}{1-k(H\alpha)/k(H\beta)}\right)
\end{equation}

where $[H\alpha/H\beta]_{obs}$ is the observed line ratio and
$k(\lambda)$ is the extinction law. We assume a foreground MW
extinction law as parameterized by \cite{pei92}. SMC and LMC laws were
also tested \citep[also using the][parametrization]{pei92}, and no
significant change was observed in the deduced extinction values
(these 3 extinction laws are practically identical at these
wavelengths). To correct for Galactic extinction towards NGC 5194 we
use a value of $A_B=0.152$, taken from \cite{schlegel98}. 

In order to test the reliability of our Balmer decrement extinction
values, we compare our corrected H$\alpha$ fluxes to
corrected Pa$\alpha$ fluxes. The
hydrogen recombination Pa$\alpha$ line at 1.87$\mu$m, although one
order of magnitude fainter than H$\alpha$, is very weakly absorbed by
dust, and hence provides an unbiased estimate of the intrinsic SFR even
in highly extincted regions \citep{scoville01}.  Most recent studies of
spatially resolved star formation in nearby disk galaxies
use recipes to account for dust obscured star formation which are
ultimately linked to a calibration based on Pa$\alpha$
\citep{calzetti05, kennicutt07, bigiel08, leroy08}. In particular,
\cite{calzetti05} finds a tight linear correlation between the
24$\mu$m luminosity of star forming regions in NGC 5194 and their
P$\alpha$ luminosities, providing justification for the use of linear
combinations of 24$\mu$m fluxes with either H$\alpha$ or UV fluxes to
estimate the intrinsic SFR in the other mentioned works. In our case,
if the extinction corrected H$\alpha$ fluxes linearly correlate with the
corrected Pa$\alpha$ fluxes, following the intrinsic line
ratio expected from recombination theory, then we can confirm that our
extinction values have been properly estimated. In that case we
can do without the IR data, and apply an extinction
correction to the measured H$\alpha$ fluxes which is solely based in
the optical spectra.

\begin{figure}[t]
\begin{center}
\epsscale{1.20}
\plotone{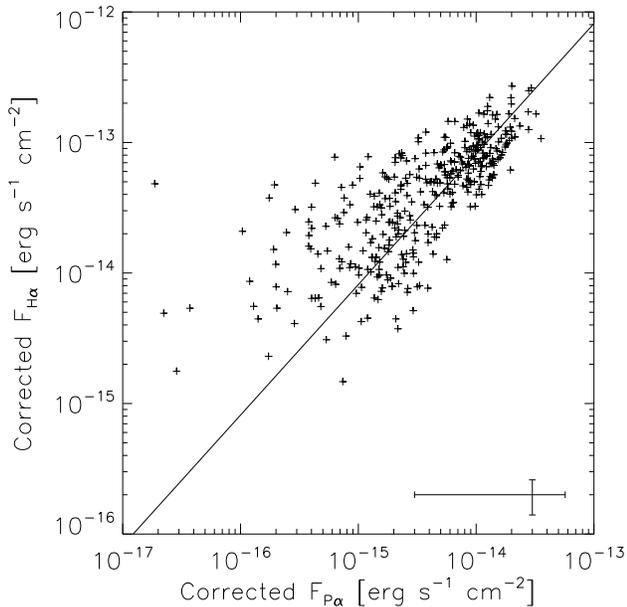}
\caption{H$\alpha$ versus Pa$\alpha$ fluxes of all regions showing
  5$\sigma$ detections of Pa$\alpha$ emission in the NICMOS
  narrow-band image. Fluxes are corrected for dust extinction using
  the Balmer decrement derived values. The solid line shows the
  H$\alpha$/Pa$\alpha$=8.15 ratio predicted by
  recombination theory. Median error bars for the corrected fluxes are shown.}
\label{fig-5}
\end{center}
\end{figure}

We measure P$\alpha$ fluxes for all 735 regions in the NICMOS F187N
continuum subtracted narrow-band image (see \S 4.1), using
apertures matching the size of the VIRUS-P fibers.
Figure \ref{fig-5} shows extinction corrected Pa$\alpha$ versus
H$\alpha$ fluxes for all regions showing
5$\sigma$ detections of Pa$\alpha$ emission in the NICMOS
narrow-band image. Both lines have been corrected using the Balmer
decrement derived extinction, and a MW extinction law. The solid line
in Figure \ref{fig-5} shows the theoretical H$\alpha$/Pa$\alpha$=8.15
ratio taken from \cite{osterbrock06}. The
observed line ratios are in agreement with the theoretical value, 
and the scatter can be attributed mostly to measurement
errors. This confirms that H$\alpha$ fluxes, once corrected for dust
obscuration using the Balmer decrement derived extinction, can provide
an unbiased measure of the intrinsic SFR in the disks of normal face-on
spirals.

\section{Measurement of Gas Mass Surface Densities}

\def\arcsec{^{\prime\prime}}
\def\hi{H{\sc i}}
\def\co1{CO~($J$=1--0)}

In order to measure the atomic and molecular gas surface density at
the position of each of the 735 regions under study, we measure
integrated intensities in the THINGS 21 cm and the BIMA SONGS CO J=1-0
maps, and translate them into gas surface densities using the
calibrations presented below. The intensities are measured over an
area equal to the beam size of each map. At each of the 735 fiber 
positions we perform aperture photometry on
the 21 cm and CO maps, and measure the integrated gas intensity in
apertures of effective radius $r_{\rm eff}$=$\sqrt{ab}/2$, where $a$ and
$b$ are the major and minor axis of the beam of each map. This
translates in an effective apperture diameter of 5.7$''$ and 5.4$''$
for the 21 cm and CO maps respectively, which is well matched to the
VIRUS-P spatial resolution which is set by the convolution of a
4.3$''$ diameter fiber and a 2$''$ FWHM seeing disk.

To convert the 21 cm intensities in atomic hydrogen column densities 
we use the following relation adapted from \cite{walter08},

\begin{equation} 
N_{HI}=1.823\times 10^{18}\left(\frac{T_B}{\rm K\; km\; s^{-1}
  \;sr}\right) {\rm cm}^{-2}
\end{equation}

where $T_{B}$ is the velocity integrated surface brightness 
temperature in the 21 cm map. To convert the CO J=1-0 intensities 
to H$_2$ column densities we use the CO to H$_2$ conversion 
factor $X_{CO}$ from \cite{bloemen86} so,

\begin{equation} 
N_{H_2}=2.8\times 10^{20}\left(\frac{T_b}{\rm K\; km\; s^{-1}
  \;sr}\right) {\rm cm}^{-2}
\end{equation}

where $T_{B}$ is the velocity integrated surface brightness 
temperature in the CO J=1-0 map. The $X_{CO}$ factor used here was
chosen for consistency with \cite{kennicutt07}, and differs from the
$X_{CO}=2.0 \times 10^{20} (K km s^{-1})^{-1}$ factor used by
\cite{bigiel08}. Current uncertainties in $X_{CO}$ are of the order of
a factor of 2, and the true value depends on assumptions about the dynamical
state of GMCs \citep{blitz07}. In any case, using a different $X_{CO}$
can only introduce an offset in the normalization of the SFL and
should not change its observed shape.

Finally, the atomic and molecular gas surface densities are derived
from the column densities using the following relations,

\begin{equation}
\Sigma_{HI}=m_H N_{HI} \cos{i}
\end{equation}

\begin{equation}
\Sigma_{H_2}=2 m_H N_{H_2} \cos{i}
\end{equation}

where $m_H$ is the hydrogen atom mass and $i=20^{\circ}$ is the
inclination of NGC 5194 as measured by \cite{tully74}. These
correspond to hydrogen gas surface densities, and should be multiplied
by a factor $\sim$1.36 to account for the mass contribution of helium
and heavier elements.

\begin{figure}[b]
\begin{center}
\epsscale{1.2}
\plotone{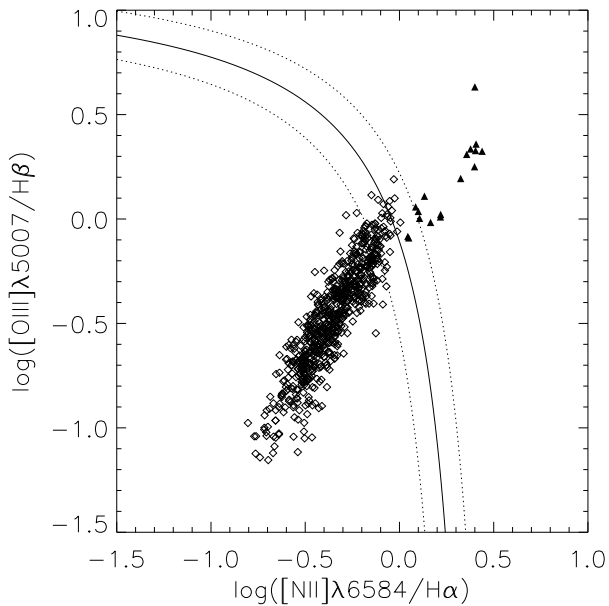}
\caption{[NII]$\lambda$6584/H$\alpha$ versus
[OIII]$\lambda$5007/H$\beta$ line ratio for the 735 regions.
The solid line marks the theoretical threshold of \cite{kewley01} separating
AGNs from star-forming galaxies. Dotted lines mark
the $\pm$0.1 dex uncertainty in the threshold modeling. The 17 regions above the
threshold and having angular distances to the galaxy nucleus of
$<15''$ are flagged as ``AGN affected'' and are shown as filled
triangles. Open diamonds show the 718 regions unaffected by AGN contamination used
to construct the SFL.}
\label{fig-6}
\end{center}
\end{figure}

\section{Photoionization and shock-heating by the central AGN}

The center of NGC 5194 hosts a weak active nucleus. The emission-line
ratios in the narrow-line region around the AGN are
consistent with those of typical Seyfert nuclei \citep[and
references therein]{bradley04}. X-ray {\it Chandra} observations show the nucleus
and two extended emission components extending $\sim 15''$ North and
$\sim 7''$ South of it \citep{terashima01}. Bipolar extended radio emission
spatially coincident with the X-ray emission, as well as weak jet with
a position angle of 158$^{\circ}$ connecting the nucleus with the
southern radio lobe was observed by \cite{crane92} and further
confirmed by \cite{bradley04}. All the observations are consistent
with the gas in the inner nuclear region ($r<1''$) being dominantly
photoionized by the central AGN, and the outer parts showing extended
emission, arising from shock-heating by a bipolar outflow.

For the purpose of constructing the SFL, we want
to exclude regions whose main source of ionization is not UV flux coming
from massive star-formation. Regions in which the gas is photoionized
by the AGN or shock-heated by the jet will emit in H$\alpha$ and
mimic star-formation. 

In order to identify these regions we use emission-line ratio
diagnostics commonly used to distinguish normal from active galaxies
\citep{veilleux87, kewley01}. Figure \ref{fig-6} shows the extinction
corrected [NII]$\lambda$6584/H$\alpha$ versus
[OIII]$\lambda$5007/H$\beta$ line ratios for all the regions. The
solid and dotted lines mark the theoretical threshold separating
AGNs from star-forming galaxies proposed by \cite{kewley01} and the
$\pm$0.1 dex uncertainty in their modeling. To avoid the
rejection of regions unaffected by AGN contamination which scatter above the 
threshold, we impose a double
criteria. We flag as ``AGN affected'', all the region lying above
the threshold, and at an angular distance of less than $15''$
(600 pc) from the nucleus of the galaxy. Filled triangles in Figure
\ref{fig-6} correspond to the 17 regions complying with both criteria. Open
diamonds correspond then to the 718 regions unaffected by AGN contamination we will
use to construct the SFL. Notice that
none of these regions lie above the +0.1 dex model uncertainty dotted
line, and that the ones lying above the threshold seem to follow the
same sequence traced by the regions unaffected by AGN contamination below
it. These fibers showing high line ratios but not associated with the
central AGN fall in the inter-arm regions of the galaxy, and have a
spectrum that is dominated by the DIG (\S 8).

\begin{figure}[t]
\begin{center}
\epsscale{1.2}
\plotone{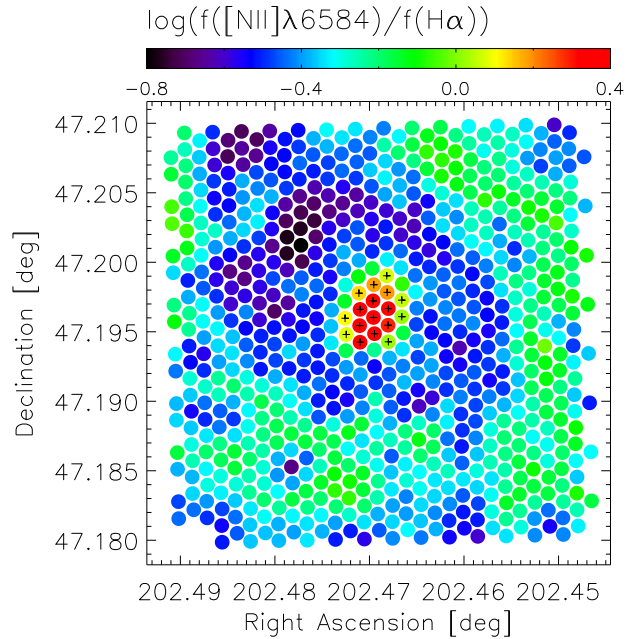}
\caption{Map of the [NII]$\lambda$6584/H$\alpha$ emission line ratio
  in the central region of NGC 5194. Regions flagged as ``AGN
  affected'' are marked by black crosses.}
\label{fig-7}
\end{center}
\end{figure}

Figure \ref{fig-7} shows a map of the [NII]$\lambda$6584/H$\alpha$
line ratio. Regions flagged as ``AGN affected'' are marked with black
crosses. It can be seen that they have high emission-line ratios
typical of AGN, and that they fall in a region which is spatially coincident with
the extended radio and X-ray emission observed around the nuclei. The
``AGN affected'' region is elongated in a similar direction to the
measured PA=158$^{\circ}$ of the radio jet \citep{crane92, bradley04,
  terashima01}. Figure \ref{fig-7} clearly shows the enhanced
line ratio in the inter-arm regions of NGC 5194. These high
ratios originate in the DIG of the galaxy and are discussed in the
following Section.

\section{Contribution from the Diffuse Ionized Gas and Calculation of SFR Surface Densities}

If we were to calculate $\Sigma_{SFR}$ using the extinction corrected
H$\alpha$ flux observed on each region, we would be working under the
assumption that all the emission observed in a given line of sight
towards the galaxy has an origin
associated with ionizing flux coming from localized star-formation in
the same region. This is not necessarily true in the presence of a
diffuse ionized component in the ISM of the galaxy. The role of the
diffuse ionized gas (DIG, a.k.a. warm ionized medium, WIM) as an
important component of the ISM of star-forming disk galaxies in the
local universe has been properly established during the last two
decades (e.g. see reviews by \citealt{mathis00} and
\citealt{haffner09}). The existence of a significant component of
extra-planar ionized hydrogen in a galaxy requires that a fraction of
the ionizing Lyman continuum photons generated in star forming regions
in the disk escapes and travels large distances of the order of
kiloparsecs before ionizing neutral hydrogen at large heights above
the disk. These distances are one
order of magnitude larger than the Str${\rm \ddot{o}}$mgren radii associated with the
most massive O stars, and the ionizing flux is thought to escape
through super-bubbles in a complex hydrogen density and ionization
distribution created by supernovae, stellar winds, and large scale
ionization by OB associations (e.g. \citealt{dove00}). 

Under these conditions a hydrogen atom emitting an H$\alpha$ photon
observed to come in the direction of a certain region of the galaxy is not
necessarily required to have been ionized by locally produced UV
photons in the same region. Hence the H$\alpha$ flux measured in each
region is the sum of the flux coming from locally star-forming H II
regions in the disk, and a contribution from the DIG. In order to
properly estimate $\Sigma_{SFR}$ and the spatially resolved SFL we
need to separate and subtract the DIG contribution from the observed
H$\alpha$ fluxes.

\begin{figure}[t]
\begin{center}
\epsscale{1.2}
\plotone{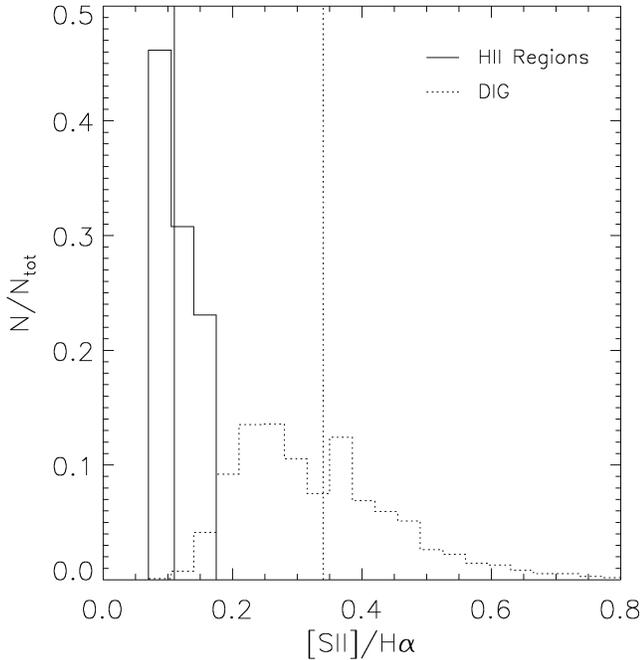}
\caption{Histogram of the [SII]/H$\alpha$ of H II regions (solid) and
  pointings towards DIG (dotted) in the Milky Way as measured by WHAM
  \citep{madsen06}. Vertical lines mark the mean values for the two distributions.}
\label{fig-8}
\end{center}
\end{figure}

Low-ionization line ratios like [NII]$\lambda$6584/H$\alpha$ and
[SII]$\lambda$6717/H$\alpha$ (hereafter [SII]/H$\alpha$)
are observed to be greatly enhanced in the DIG, as compared to the
typical values observed in H II regions \citep{reynolds85,
  hoopes03}. Recent results from The Wisconsin H$\alpha$ Mapper
(WHAM\footnote{http://www.astro.wisc.edu/wham/}) 
sky survey by \cite{madsen06} show that H II regions in the Milky Way
have a typical ([SII]/H$\alpha)_{H II}$=0.11 with a small rms scatter from
region to region of only $\Delta$([SII]/H$\alpha)_{H II}$=0.03. On the
other hand, high galactic latitude pointings sampling the DIG
component show a mean ([SII]/H$\alpha)_{DIG}$=0.34, with a large scatter
from pointing to pointing of
$\Delta$([SII]/H$\alpha)_{DIG}$=0.13. Figure \ref{fig-8} shows a
histogram of the [SII]/H$\alpha$ line ratios taken from \cite{madsen06} for H II regions and the DIG
as measured by WHAM. It can be seen that the [SII]/H$\alpha$ ratio provides a very useful
tool to separate the contribution from the DIG and the disk H II
regions in our spectra. The [NII]/H$\alpha$ ratio, while still
enhanced in the DIG as can be clearly seen in Figure \ref{fig-7},
shows a much larger scatter both for H II regions 
and pointings towards the DIG, and does not provide such a clean
separation as the [SII]/H$\alpha$ ratio (see Figure 21 in
\cite{madsen06}).

We model the measured H$\alpha$ flux of each region as the sum of
a contribution from H II regions plus a contribution from the DIG, so

\begin{equation}
\begin{array}{ccl}
f(H\alpha) & = & f(H\alpha)_{H II}+f(H\alpha)_{DIG}\\\;\\
           & = & C_{H II}f(H\alpha)+C_{DIG}f(H\alpha)
\end{array}
\end{equation}

where $C_{H II}$ is the fraction of the total flux coming from local
star-forming regions in the disk, and $C_{DIG}$=(1-$C_{H II}$). The
observed [SII]/H$\alpha$ ratio is then given by,

\begin{figure}[b]
\begin{center}
\epsscale{1.10}
\plotone{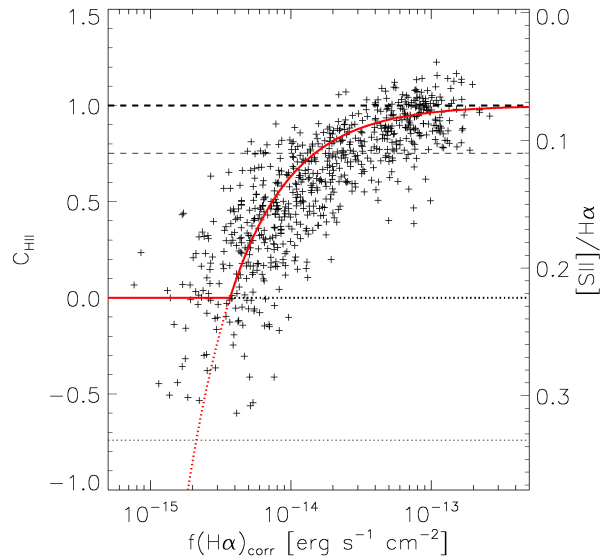}
\caption{Observed [SII]/H$\alpha$ emission line ratio for the 718
  regions unaffected by AGN contamination. The thin dashed and dotted lines show
  the mean ratio observed in H II regions and pointings towards the DIG
in the Milky Way respectively. The thick dashed and dotted lines show
the former ratios scaled down by a factor $Z'=1.0/1.5$. The left axis
shows the fraction of the flux coming from H II regions in the disk
given by Equation 8. The solid red curve shows the DIG correction
applied to the data given by Equation 9, and the continuation of the
function to fluxes lower than $f_0$ is marked by the dashed red line.}
\label{fig-9}
\end{center}
\end{figure}

\begin{equation}
\frac{[SII]}{H\alpha}=Z'\left[C_{H II}\left(\frac{[SII]}{H\alpha}\right)_{H II}+C_{DIG}\left(\frac{[SII]}{H\alpha}\right)_{DIG}\right]
\end{equation}

\begin{figure*}[t]
\begin{center}
\epsscale{1.15}
\plotone{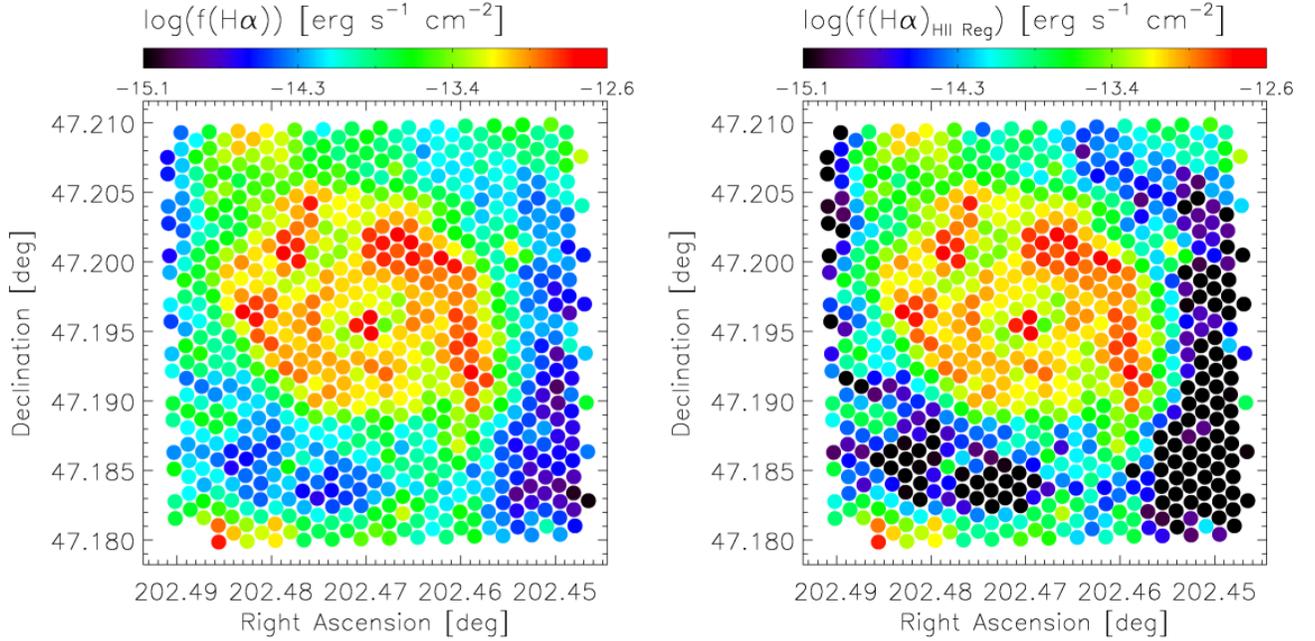}
\caption{{\it Left:} Map of the extinction corrected H$\alpha$ nebular
emission flux in the central 4.1$\times$4.1 kpc$^2$ of NGC 5194. {\it
  Right:} Same map after removing the DIG contribution to 
the H$\alpha$ emission line flux, that is, showing only the flux
coming from H II regions in the disk of NGC 5194.}
\label{fig-10}
\end{center}
\end{figure*}

where $Z'=Z/Z_{MW}$ is the metallicity of NGC 5194 normalized to the Milky
Way value. Figure \ref{fig-9} shows the observed
[SII]/H$\alpha$ ratio as a function of extinction corrected H$\alpha$
flux. The left axis shows $C_{H II}$ calculated assuming
a value of $Z'=1.0/1.5$. \cite{bresolin04} measured the oxygen and
sulfur abundance gradient in NGC 5194 using multi-object spectroscopy
of 10 H II regions spanning a large range in radii. Integrating his
best-fitted oxygen abundance gradient out to a radius of 4.1 kpc provides an mean
12+log(O/H)=8.68, which is 1.55 times lower than the solar oxygen
abundance measured by \cite{grevesse96}. Although a large scatter is
observed in the literature for both the solar oxygen abundance and the
oxygen abundance in Milky Way H II regions \citep{grevesse96,
allendeprieto01, shaver83, deharveng00}, it can be seen in Figure 
\ref{fig-9} that using a factor of 1.5 implies that the brightest
H$\alpha$ emitting regions in NGC 5194 are completely dominated by
emission from H II regions in the disk, having $C_{H II}\sim 1$ with a
scatter that is consistent with the intrinsic scatter of 0.03 measured in the
Milky Way by \cite{madsen06}. These brightest regions trace the spiral
structure of the galaxy and are expected to be H II region dominated 
since on high star-formation regions the disk should outshine the DIG 
by many orders of magnitude.

\begin{figure*}[t]
\begin{center}
\plotone{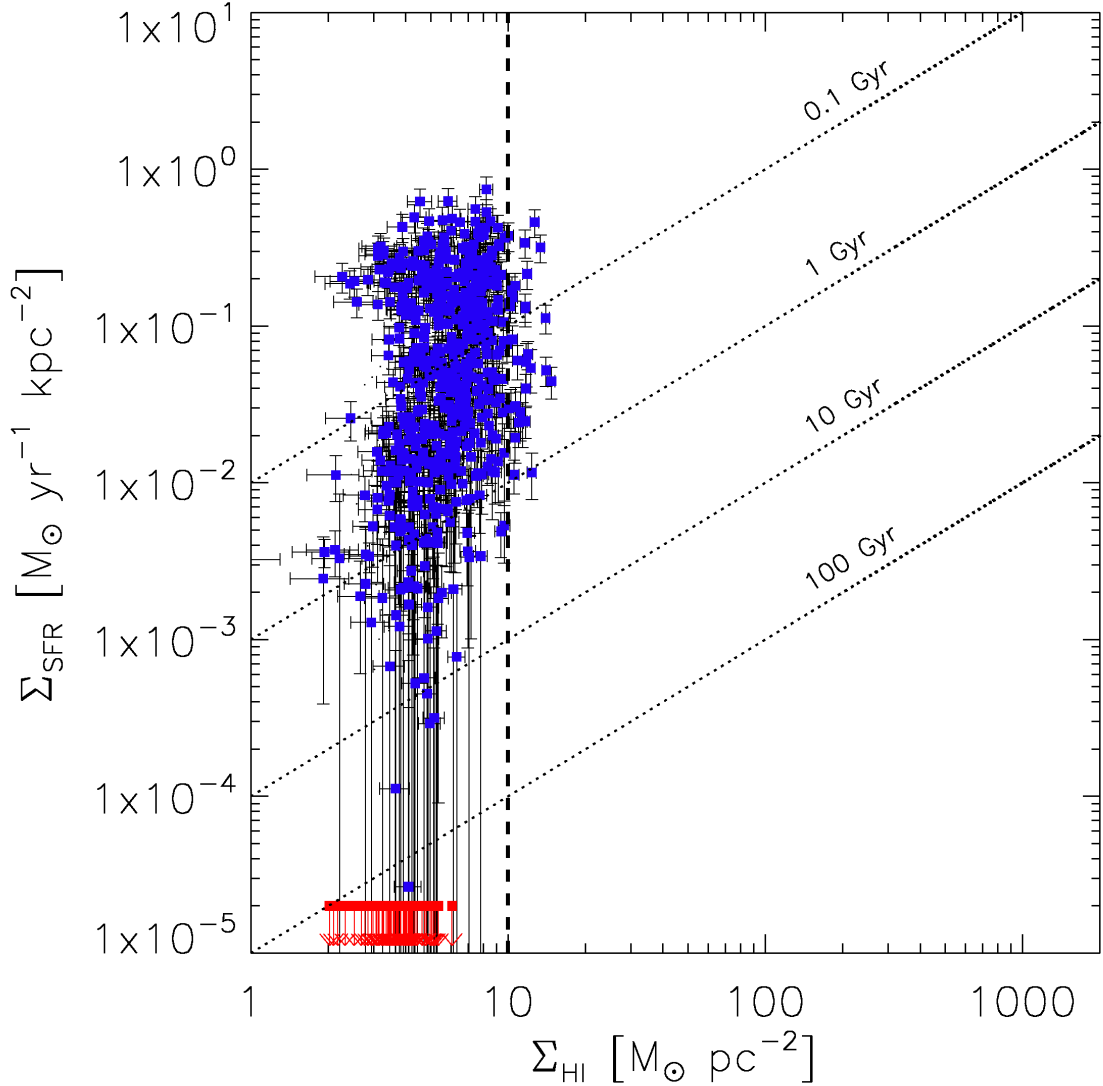}
\caption{Atomic gas suface density versus SFR surface density for the
718 regions unaffected by AGN contamination. Upper limits in $\Sigma_{SFR}$
correspond to regions with $C_{H II}=0$. The vertical dashed line marks
the HI to H$_2$ transition threshold at 10 M$_{\odot}$pc$^{-2}$. The
diagonal dotted lines correspond to constant depletion timescales
$\tau={\rm SFE}^{-1}$ of 0.1, 1, 10 and 100 Gyr.}
\label{fig-11}
\end{center}
\end{figure*}

There is a clear correlation between $C_{H II}$ and the H$\alpha$
flux. The observed trend is consistent
with the DIG dominating the spectrum of fainter H$\alpha$ regions, and
the H II regions in the disk outshining the DIG in the brightest ones.
The scatter is large mostly because of intrinsic scatter in the line ratio (see
Figure \ref{fig-8}). In order to compute a robust DIG correction, we 
fit the $C_{H II}$ values using the simple functional form,

\begin{equation}
C_{H II}=1.0-\frac{f_0}{f(H\alpha)}\; ;\;\;({\rm for}\;f(H\alpha)>f_0)
\end{equation}

where $f_0=3.69\times 10^{-15}$ erg s$^{-1}$cm$^{-2}$ is the flux at
which the DIG contributes 100\% of the 
emission, and hence $C_{H II}=0$ for $f(H\alpha)\leq f_0$. The
correction is shown as the red solid line in Figure \ref{fig-9}. We
multiply the extinction corrected H$\alpha$ fluxes by the above
correction factor in order to remove any contribution from the DIG in
NGC 5194. It is worth noting that using Equation 9 to remove the DIG
is equivalent to subtracting a constant DIG flux value $f_0$ for all
regions with $f(H\alpha)>f_0$ (the large majority of the regions).
Hence, the line ratio distribution presented in Figure \ref{fig-9} is
very well fitted by a flat DIG component.

Figure \ref{fig-10} presents maps of the extinction corrected
H$\alpha$ emission line flux before and after the DIG correction is
applied. It can be seen clearly how the H$\alpha$ emission traces the
spiral pattern of star-formation. The correction leaves the
H$\alpha$ flux coming from the brightest star-forming regions
practically unchanged, while removing the contribution from the DIG
which dominates the observed spectrum in the inter-arm regions of the
galaxy. The latter can also be appreciated in Figure \ref{fig-7}, which
shows an enhanced [NII]/H$\alpha$ ratio typical of the DIG in the
inter-arm regions, and normal H II region ratios throughout the spiral
arms.

Integrating over the complete observed area, the DIG contributes only
11\% of the total H$\alpha$ flux. Previous photometric measurements of the
diffuse ionized fraction in nearby spiral galaxies, including NGC 5194, yield median
diffuse fractions of $\sim$50\% \citep[e.g.][]{ferguson96, hoopes96,
greenawalt98, thilker02, oey07}. These studies are performed either by 
masking of H II regions or by discrete H II region photometry in
H$\alpha$ narrow-band images. Although it will be seen in \S 10 that the assumption of a
constant [NII]/H$\alpha$ ratio throughout the galaxy used to correct
the narrow-band images in all the above studies can introduce
overestimations of the DIG H$\alpha$ brightness of up to 40\%, this
effect is small, and cannot account for the difference between our diffuse
fraction and the typical values found in the literature. The
difference is most likely due to the fact that our observations are
limited to the highly molecular, and hence strongly star-forming
central part of the galaxy. Our measured diffuse ionized fraction is
then only a lower limit to the DIG contribution over the whole galaxy, since at larger
radii the relative contribution from H II regions is expected to
significantly decrease. 
Though the DIG contribution to the integrated H$\alpha$ luminosity in
the central region of NGC 5194
could be small, on the small scales sampled by the VIRUS-P fibers the
DIG can account for 100\% of the observed H$\alpha$ flux, especially
in between the spiral arms where H II regions are
rare. Given the clear dependence of the above correction with
H$\alpha$ flux, failing to correct for this effect introduces a bias
in the SFL towards shallower slopes.

The corrected H$\alpha$ emission-line fluxes are transformed
into H$\alpha$ luminosities using the assumed distance to NGC 5194 of
8.2 Mpc. Since the DIG is suspected to arise from UV photons escaping
star forming regions in the disk, not accounting for these photons
should introduces a systematic underestimation of the SFR. The
challenge resides in our inability to tell from where in the disk
these UV photons come from. To ameliorate this problem, we scale the
H$\alpha$ luminosities by a factor of 1.11, which is equivalent to
assuming that the UV photons ionizing the DIG were originated in the
star-forming regions in the disk proportionally to their intrinsic UV
luminosities. These scaled luminosities ($\rm L_{corr}(\rm{H\alpha})$) are used to estimate the
SFR for each of the 718 regions. We use the
calibration presented in \cite{kennicutt98a}, for which the SFR is
given by,

\begin{equation}
\rm{SFR}\;[M_{\odot}\rm{yr}^{-1}]=7.9\times10^{-42}L_{corr}(\rm{H\alpha})\;[\rm{erg\;s^{-1}}]
\end{equation}

The above calibration assumes a Salpeter IMF over the range of stellar
masses 0.1-100 $M_{\odot}$. To convert to the Kroupa-type
two-component IMF used in \cite{bigiel08}, the SFR must be multiplied
by a factor of 0.63.

The SFRs for individual regions are then converted to SFR surface densities
($\Sigma_{SFR}$). Following \cite{kennicutt07}, we divide the SFR by
the projected area on the sky of the $4.3''$ (172 pc) diameter regions
sampled by each fiber on the IFU, and multiply it by a factor of
$\rm{cos}(20^{\circ})$ to account for the inclination of NGC 5194
\citep{tully74}.

\section{The Spatially Resolved Star Formation Law}

\begin{figure*}[t]
\begin{center}
\plotone{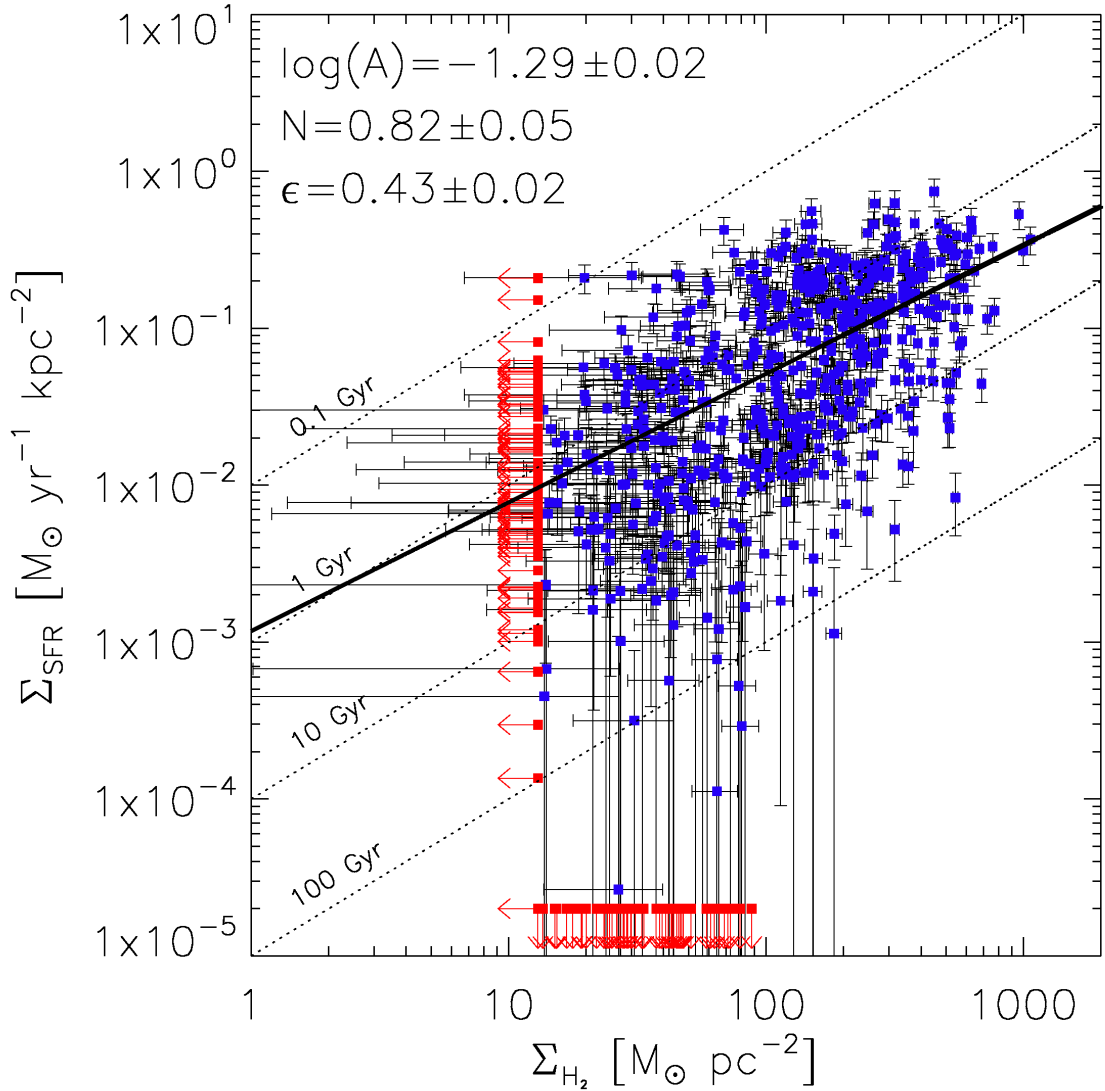}
\caption{Molecular gas suface density versus SFR surface density for the
718 regions unaffected by AGN contamination. Upper limits in $\Sigma_{SFR}$
correspond to regions with $C_{H II}=0$. Upper limits in $\Sigma_{H_2}$
correspond to regions with non-detection in CO at the 1$\sigma$
level.  The
diagonal dotted lines correspond to constant depletion timescales
$\tau={\rm SFE}^{-1}$ of 0.1, 1, 10 and 100 Gyr. Also shown is the best-fitted power law from the Monte Carlo
method (black solid line), and the best-fitted parameters.}
\label{fig-12}
\end{center}
\end{figure*}

\begin{figure*}[t]
\begin{center}
\plotone{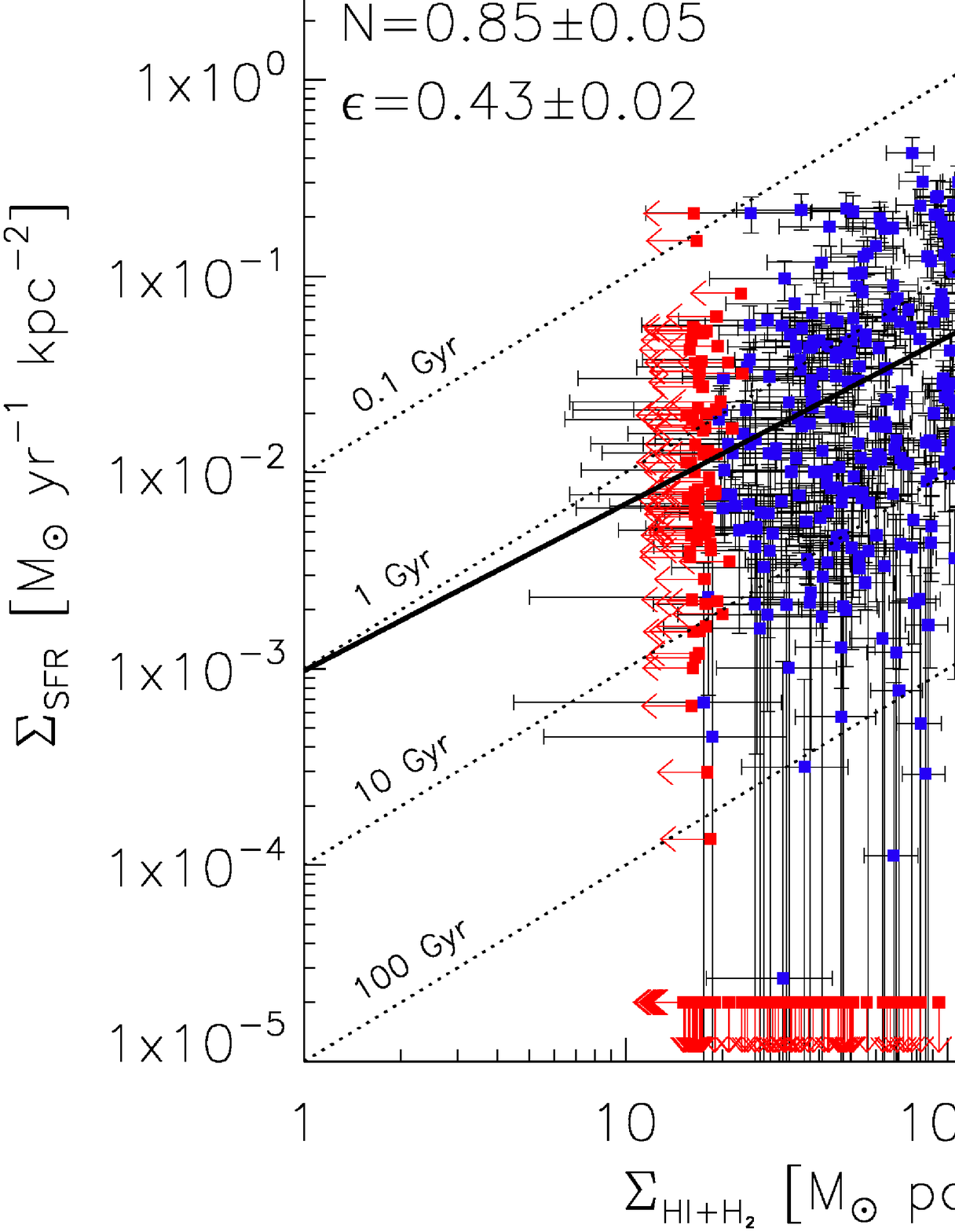}
\caption{Total gas suface density versus SFR surface density for the
718 regions unaffected by AGN contamination. Upper limits in $\Sigma_{SFR}$
correspond to regions with $C_{H II}=0$. Upper limits in $\Sigma_{HI+H_2}$
correspond to regions with non-detection in CO at the 1$\sigma$
level.  The
diagonal dotted lines correspond to constant depletion timescales
$\tau={\rm SFE}^{-1}$ of 0.1, 1, 10 and 100 Gyr. Also shown is the best-fitted power law from the Monte Carlo
method (black solid line), and the best-fitted parameters.}
\label{fig-13}
\end{center}
\end{figure*}

The observed relations between
$\Sigma_{SFR}$ and the gas surface densities of different components
of the ISM ($\Sigma_{HI}$, $\Sigma_{H_2}$, and
$\Sigma_{HI+H_2}$) are presented in Figures
\ref{fig-11}, \ref{fig-12} and \ref{fig-13}. Error bars in gas surface
densities
correspond to the 1$\sigma$ uncertainties given in \S 4.1 and \S
4.2. Error bars in the SFR surface density include a series of
uncertainties that we proceed to describe. First we consider the uncertainty in 
the observed H$\alpha$ fluxes. This comes from the fitting of the
H$\alpha$ line described in \S 5.2, which was performed considering the
observational error in the spectrum (obtained from the error maps
described in \S 3). Second, the 
uncertainty in the dust extinction correction is included by
propagating the fitting errors of the observed H$\alpha$ and H$\beta$ fluxes
through Equation 2. Finally, in order to account for the error
associated with the DIG correction, we introduce a 20\% uncertainty in
$\Sigma_{SFR}$, consistent with the median scatter of the points in
Figure \ref{fig-9} around the correction used. All these uncertainties
are summed in quadrature to account for the error bars in
$\Sigma_{SFR}$. We do not consider errors in the flux calibration
which are expected to be of $\sim$5\%, nor in the CO to H$_2$
conversion factor. The later is currently highly uncertain and might change
by up to a factor of 2 depending on assumptions about the dynamical
state of GMCs \citep{blitz07}. In any case, these two sources of
systematic errors enter the SFL as multiplicative factors. Hence, they can
only introduce a bias in the normalization of the SFL, and should not
affect the fitted values of the slope and the intrinsic scatter.

From Figure \ref{fig-11}
it is clear that $\Sigma_{SFR}$ shows a very poor correlation with
$\Sigma_{HI}$, since regions having similar atomic gas budgets can have
star formation activities that differ by more than 3 orders of
magnitude. We observe an evident saturation in the atomic gas surface
density at $\Sigma_{HI}\approx 10\;{\rm M_{\odot}pc^{-2}}$. Also, there
is a slight inversion in the sense of the correlation at high
$\Sigma_{SFR}$, associated with the central part of the galaxy due to
the presence of a minimum in the HI profile \citep{bigiel08}. These
HI ``holes'' are common in the centers of spiral galaxies, and in them
the ISM is fully dominated by molecular hydrogen while the atomic gas
is almost completely depleted. The saturation at $10\;{\rm M_{\odot}pc^{-2}}$
has been previously observed by \cite{wong02} using azimuthally
averaged data, and further confirmed to be a widespread phenomena in
normal spirals by \cite{bigiel08} using 2D spatially resolved
measurements. It is thought to be related to a threshold in
surface density at which a phase-transition from atomic to molecular
gas occurs in the ISM \citep{krumholz09a}. Given the lack of
correlation between $\Sigma_{HI}$ and
$\Sigma_{SFR}$, we do not attempt to fit a atomic gas SFL. We
restrict our analysis to the modeling of the molecular and total gas
correlations with the star-formation activity. These correlations are
usually well described by a power-law function\citep{schmidt59,
kennicutt98b}. 

It has been established that the observed rms
dispersion about a power-law 
in these SFLs is much larger than the observational
uncertainties \citep{kennicutt98b, kennicutt07}, implying the
existence of significant intrinsic scatter of physical origin in the
relations. 
However, previous works have not introduced this
intrinsic scatter into the parameterization of the SFL,
and authors restrict themselves to measure the scatter after
fitting a power-law to the data. In this work, we incorporate the
intrinsic scatter in the SFL, which we
parameterize as:

\begin{equation}
\frac{\Sigma_{SFR}}{1{\rm M_{\odot}yr^{-1}kpc^{-2}}} =
A\left(\frac{\Sigma_{gas}}{100{\rm M_{\odot}pc^{-2}}}\right)^N \times 10^{\;{\cal N}(0,\epsilon)}
\end{equation}

where $A$ is the normalization factor, $N$ is the slope, and ${\cal
 N}(0,\epsilon)$ is a logarithmic deviation from the power-law, drawn
from a normal distribution with zero mean and standard deviation
$\epsilon$. The value of $\epsilon$ corresponds to the intrinsic
scatter of the SFL in logarithmic space. The factor
$10^{\;{\cal N}(0,\epsilon)}$ can be interpreted as changes
of physical origin in the star-formation efficiency for different 
regions. We chose a
pivot value for the normalization of
100${\rm M_{\odot}pc^{-2}}$, which is roughly at the center of the
distribution of measured $\Sigma_{gas}$ values, in order to minimize
the covariance between the slope and the normalization. When
comparing the normalization factors derived here with other fits
found in the literature, this must be taken into account. Most works
quote normalizations at $1{\rm M_{\odot}pc^{-2}}$, while
\cite{bigiel08} quotes normalizations at $10{\rm M_{\odot}pc^{-2}}$.

Previous measurements of the spatially resolved SFL
use different algorithms to fit a power-law to the data. Usually a
linear regression in logarithmic space is performed, but methods
differ in the treatment of error bars. \cite{kennicutt07} used a
FITEXY algorithm \citep{numrec}, which has the advantage of
incorporating errors in both the ordinate and abscissa coordinates,
although errors must be assumed to be symmetric in logarithmic space,
which is not always the case. 
\cite{bigiel08} used an ordinary
least-squares (OLS) bisector method \citep{isobe90} giving the same
weighting to every data point. Both methods have the
disadvantage of not being able to incorporate upper limits in the
minimization. Our data is mainly limited by the
sensitivity of the CO intensity maps as can be seen in Figure
\ref{fig-12}, where 93 of the 718 regions unaffected by AGN contamination are undetected in
CO and hence we can only provide upper limits for their molecular gas
surface densities. This is also the case in the works mentioned
above. As will be seen in \S 11, these upper limits contain important
information regarding the slope of the spatially resolved SFL, 
and neglecting them biases the fits towards steeper slopes. We
introduce and use a new method for fitting the SFL which is not
affected by the above issues.

\begin{figure*}[t]
\begin{center}
\epsscale{1.2}
\plotone{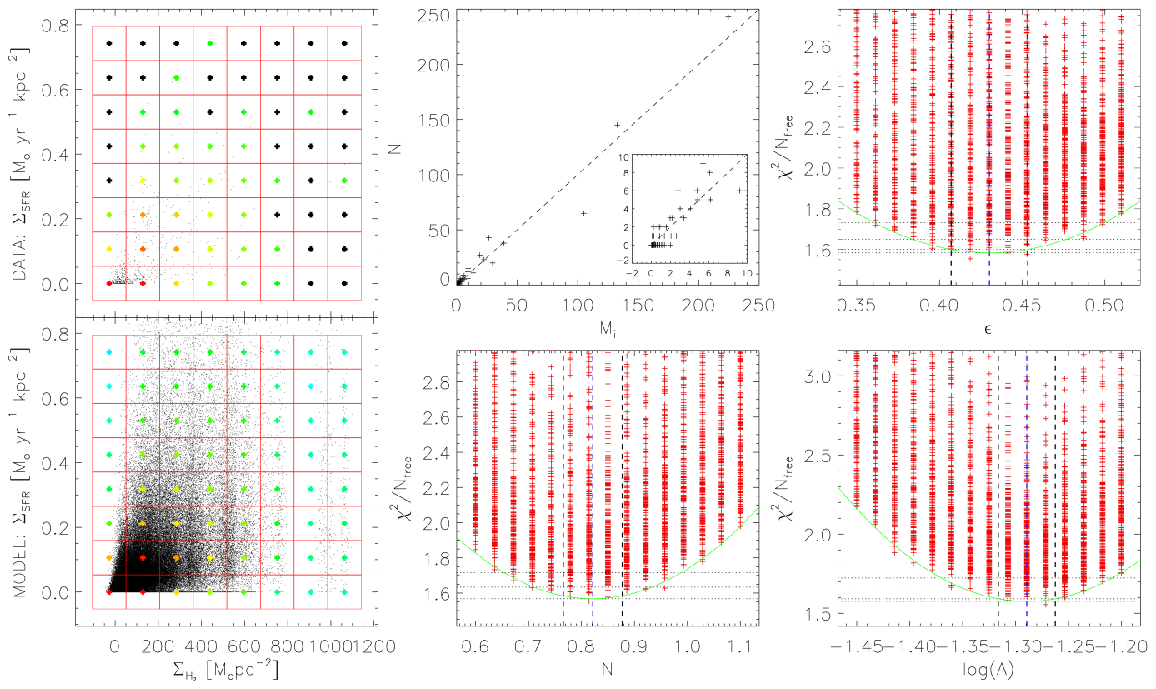}
\caption{{\it Left:} The observed molecular SFL in linear space (top),
together with the 200 Monte Carlo realizations of the data for the
best-fitted parameters (bottom). The grid used to compare the model to the data is
shown in red, and each box in the grid shows a cross, color-coded
according to the number of points in the grid (with red corresponding
to the highest value and black corresponding to zero). {\it
  Center-Top:} Number of data-points per grid elements in the model
versus the data. {\it Center-Bottom and Right:} Reduced $\chi^2$ for
each of the three free parameter in the fit ($A$, $N$, and $\epsilon$), marginalized over the
other two parameters. Red crosses show the $\chi^2$ obtained for each
sampled combination of parameters. The best-fitted quadratic function to
the minimum $\chi^2$ is shown in green. The best-fitted $\chi^2$, together
with the 1$\sigma$, 2$\sigma$, and 3$\sigma$ levels are shown as horizontal dotted
lines. The blue and black vertical dashed lines marks the best-fitted
parameter and its 1$\sigma$ uncertainty respectively.}
\label{fig-14}
\end{center}
\end{figure*}

\subsection{The Fitting Method}

To fit our data we use a Monte Carlo (MC) approach combined with
two-dimensional distribution comparison techniques
commonly used in color-magnitude diagram (CMD) fitting
\citep{dolphin01}. Our method allows us to include the regions not
detected in the CO map (including the ones with negative measured
fluxes), incorporate the intrinsic scatter in the SFL
as a free parameter, and perform the fitting in linear space, avoiding
the assumption of log-normal symmetric errors. In the following, we
describe our fitting method. 

For any given set of parameters $\{A, N, \epsilon\}$, we generate 200
Monte Carlo realizations of the data. To create each realization, we take
the measured values of $\Sigma_{gas}$ as the true values and calculate the
corresponding true $\Sigma_{SFR}$ using Equation 11, drawing a new
value from ${\cal N}(0,\epsilon)$ for each point in order to introduce
the intrinsic scatter. Regions for which we measure negative CO
fluxes are assumed to have
$\Sigma_{gas}=\Sigma_{SFR}=0$. In order to account for observational
errors, data points are then offset
in $\Sigma_{SFR}$ and $\Sigma_{gas}$ by random quantities given by the
observed measurement error for each data point. The uncertainty in
$\Sigma_{SFR}$ is largely dominated by the errors introduced in the
dust extinction and DIG corrections. Since both corrections are
multiplicative, we apply the random offsets as multiplicative
factor drawn from a ${\cal
  N}(1,\sigma(\Sigma_{SFR})/\Sigma_{SFR})$ distribution. On the
other hand, the
error in $\Sigma_{gas}$ is dominated by systematic offsets introduced
during the combination and calibration of interferometric
data. Accordingly, the random offsets in $\Sigma_{gas}$ are
introduced in an additive manner, using values drawn from a ${\cal
  N}(0,\sigma(\Sigma_{gas}))$ distribution. It is important to notice
that while for plotting purposes, Figures \ref{fig-11}, \ref{fig-12},
and \ref{fig-13} show upper limits in $\Sigma_{gas}$ and
$\Sigma_{SFR}$, in the fitting procedure the measured values of these
data-points are used together with their usually large error bars.

Having the observed data points and the large collection of
realizations of the data coming from the model, we need to
compare the distribution of points in the
$\Sigma_{gas}$-$\Sigma_{SFR}$ plane in order to assess how well the
model fits the data given the assumed parameters. To do so, we define
a grid on the $\Sigma_{gas}$-$\Sigma_{SFR}$ plane and count
the number of data points falling on each grid element both in the
data and in the 200 realizations. This method is adapted from
\cite{dolphin01}, and it is the equivalent to the construction of Hess
diagrams used in CMD fitting. The grid covers all the observed data
points, has a resolution of $\Delta\Sigma_{gas}$=156 ${\rm
  M_{\odot}pc^{-2}}$ and of $\Delta\Sigma_{SFR}$=0.11 ${\rm
  M_{\odot}yr^{-1}kpc^{-2}}$, and is shown in the left panel of Figure
\ref{fig-14}. A single extra grid element
containing all the points in the Monte Carlo realizations falling
outside the grid and
zero observed data points is also included in the calculations below.

We average the number of points in each
grid element for the 200 Monte Carlo realizations and call this ``the
model''. In order to
compare the model to the data we compute a $\chi^2$ statistic of the
following form:

\begin{equation}
\chi^2=\sum_i\frac{(N_i-M_i)^2}{M_i}
\end{equation}

Where the sum is over all the grid elements in the
$\Sigma_{gas}$-$\Sigma_{SFR}$ plane, $N_i$ is the number of observed
data points, and $M_i$ is the number of model data points in the grid
element $i$. We sample a large three dimensional grid in parameter space with a
resolution of $\Delta$log($A$)=0.018,  $\Delta N$=0.036, and $\Delta \epsilon$=0.011,
centered around our best initial guesses for the different SFLs, and
compute $\chi^2$ for every combination of parameters in the cube. 

To exemplify our method, the left panel in Figure \ref{fig-14} shows the
observed molecular SFL in linear space, together with the best-fitted Monte Carlo
model. Overlaid are all the grid elements, color-coded according with
the density of points inside each of them. The top central panel shows
the number of points in each grid element in the model versus the data
for the best-fitted model, in this plot, deviations from the dashed line
contribute to the $\chi^2$ statistic. Also shown is the $\chi^2$ for
each parameter, marginalized over the other two. The best-fitted value
for each parameter is obtained by fitting a quadratic function to the
minimum $\chi^2$ at each parameter value sampled. Uncertainties at the
1$\sigma$, 2$\sigma$, and
3$\sigma$ levels are also shown in the plots. Notice that the sampled set of
parameters showing the minimum $\chi^2$ is always within 1$\sigma$ of
the best-fitted value deduced from the quadratic function fitting.

Thorough testing of the fitting method was carried out. The number of
Monte-Carlo simulations is high enough for consecutive runs of the
algorithm on the same data to produce best-fitted values for the
parameters that show a scatter of less than 0.1$\sigma$. The best-fitted
parameters are somewhat sensitive to the chosen grid spacing in the
linear $\Sigma_{gas}$-$\Sigma_{SFR}$ plane. Fitting of artificially
generated data-sets drawn from known parameters, showed the
grid resolution we use to be the best at recovering the intrinsic parameters
with deviations from the true values of less than 0.5$\sigma$.

\subsection{Fits to the Molecular and Total Gas Star Formation Laws}

We applied our method to fit the observed SFL in both molecular gas
and total gas. The best-fitted SFLs are shown as solid lines in Figures
\ref{fig-12} and \ref{fig-13}, where the best-fitted parameters are also
reported. For the molecular gas SFL we measure a slope
$N=0.82\pm0.05$, an amplitude $A=10^{-1.29\pm0.02}$, and an intrinsic scatter
$\epsilon=0.43\pm0.02$ dex. In the central part of NCG 5194 we are
sampling a density regime in which the ISM is almost fully molecular,
hence the total gas SFL closely follows the molecular SFL and shows
very similar best-fitted parameters. For the total gas SFL we obtain a
slope $N=0.85\pm0.05$, an amplitude $A=10^{-1.31\pm0.02}$, and an 
intrinsic scatter $\epsilon=0.43\pm0.02$ dex.

Of great interest is the large intrinsic scatter observed in the
SFL. A logarithmic scatter of 0.43 dex implies that the SFR in 
regions having the same molecular gas surface density can vary roughly
by a factor of $\sim$3. This is very important to keep in mind when
using the SFL as a star-formation recipe in theoretical models of
galaxy formation and evolution. Results from this type of modeling
should be interpreted in an statistical sense, and we must always
remind ourselves that SFRs predicted for single objects can be off by
these large factors. The bottom left panel of Figure \ref{fig-14} is
an striking reminder of the limitations involved in the use of SFLs
as star-formation recipes in analytical and semi-analytical
models. The large scatter observed is indicative of the
existence of other parameters, besides the availability of molecular gas,
which are important in setting the SFR.

As will be discussed in \S 11, the fact that we measure a slightly
sub-linear SFL is consistent recent results by \cite{bigiel08} and
\cite{leroy08}, as well as with recent theoretical modeling by
\cite{krumholz09b}, but in disagreement with the significantly
super-linear molecular and total gas SFLs measured in NGC 5194 by
\cite{kennicutt07}. Our results imply depletion times for the
molecular gas of $\tau\approx2$ Gyr, which is roughly a factor of
$\sim$100 longer than the typical free fall time of GMCs
\citep{mckee99}. These low efficiencies, of the order of 1\% per
free-fall time, are observed in a large range of spatial scales and
densities in different objects. It is seen all the way from
HCN emitting clumps, infrared dark clouds, and GMCs in the Milky Way
to the molecular ISM in large scales in normal spiral galaxies and
starburst, and is consistent with models in which star-formation is
regulated by supersonic turbulence in GMCs, induced by feedback from
star-formation itself \citep{evans09, krumholz06}. 

\section{Balmer Absorption and the N[II]/H$\alpha$ Ratio, Implications for Narrow-Band Imaging}

\begin{figure}[b]
\begin{center}
\epsscale{1.2}
\plotone{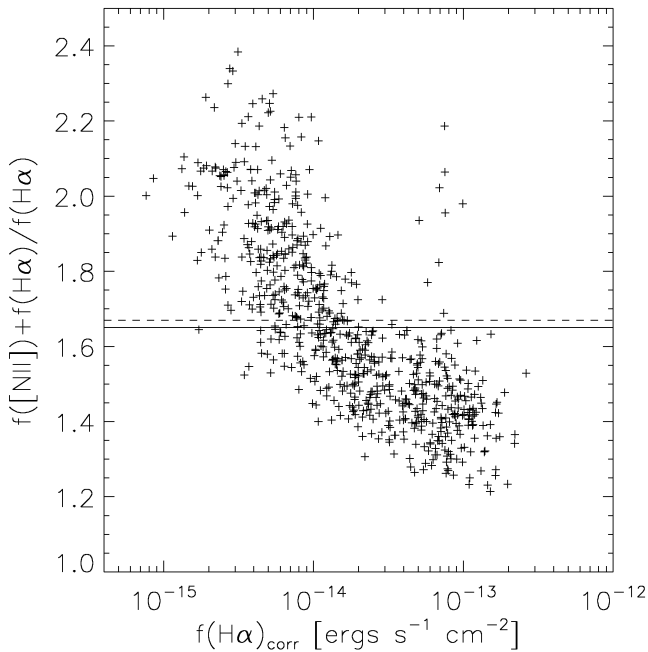}
\caption{([NII]$\lambda$6548+[NII]$\lambda$6584+H$\alpha$)/H$\alpha$
ratio as a function of extinction corrected H$\alpha$ flux for the
718 regions under study. The solid line marks the observed mean
value of 1.65. The dashed line marks the 1.67 value expected by
assuming line ratio of [NII]$\lambda$6584/H$\alpha$=0.5 and
[NII]$\lambda$6548/[NII]$\lambda$6584=0.335.}
\label{fig-15}
\end{center}
\end{figure}

Narrow-band imaging is the most widely used method for conducting
spatially resolved measurements of the H$\alpha$ emission line in
nearby galaxies. Images taken with a narrow-band filter centered at
H$\alpha$, and either a broad-band or off-line narrow-band, are
subtracted in order to remove the continuum in the on-line
bandpass. The excess flux in the on-line narrow-band is expected to
map the nebular emission. Narrow-band filters have typical FWHMs of
$\sim$70\AA , and hence suffer from contamination from the
[NII]$\lambda\lambda$,6548,6584 doublet. Also, narrow-band techniques
cannot directly separate the nebular emission from the underlying
photospheric absorption H$\alpha$. Corrections to account for these
two factors are usually applied. 

In order to correct for the underlying absorption, the continuum image
is usually scaled before subtraction so selected regions in the
galaxy, which are a priori expected to be free of H$\alpha$ emission,
show zero flux in the subtracted image. This is equivalent to
correcting for a constant H$\alpha$ absorption EW across the
galaxy (assuming that the continuum level was reliably estimated,
which might not be the case when broad-bands are used instead of
off-line narrow-bands, since the spectral slope of the stellar
continuum can vary significantly across the galaxy). The
[NII] contamination is usually taken out by assuming a
constant [NII]/H$\alpha$ ratio across the whole galaxy, which together
with the relative filter transmission at the wavelengths of the three
lines, is used to compute a correction factor which is used to scale
down the observed continuum subtracted narrow-band fluxes in order to
remove the [NII] contribution. Integral-field spectroscopy is free of
these two effects, since both the [NII] lines and the photospheric
H$\alpha$ absorption can be clearly separated from the H$\alpha$
emission (see Figure \ref{fig-3}). Thus, our observations provide an
important check on the validity of the corrections typically applied in
narrow-band studies, and the biases introduced by them.

For the [NII] correction, line ratios of
[NII]$\lambda$6584/H$\alpha$=0.5 and
[NII]$\lambda$6548/[NII]$\lambda$6584=0.335 are typically assumed
\citep{calzetti05}. Based on these ratios, a perfect H$\alpha$ filter
(i.e. one with a constant transmission across the three lines) would
measure a flux that is a factor of 1.67 higher than the H$\alpha$
flux. Figure \ref{fig-15} shows the
([NII]$\lambda$6548+[NII]$\lambda$6584+H$\alpha$)/H$\alpha$ ratio as a
function of the extinction corrected H$\alpha$ flux, as measured in
the VIRUS-P spectra of all 718 star-forming regions. Although we
measure a mean value of 1.65 (solid line), in good agreement with the
predictions from the above line ratios (dashed line), it can be seen
that the correction factor is a strong function of H$\alpha$ flux. The
fact that we observe an increasing [NII]/H$\alpha$ ratio as we go to
fainter H$\alpha$ fluxes is consistent with the nebular emission in the
faintest parts of the galaxy (mainly the inter-arm regions) being
dominated by the DIG component of the
ISM (see Figure \ref{fig-7} and \S 8).

\begin{figure}[t]
\begin{center}
\epsscale{1.15}
\plotone{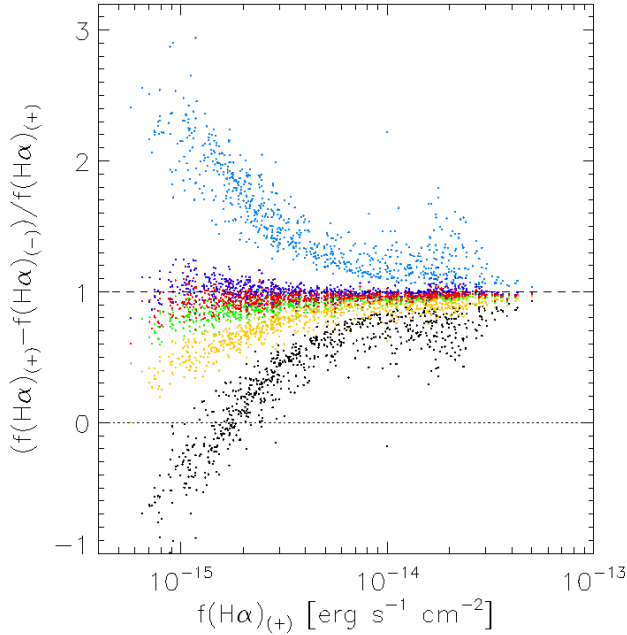}
\caption{Bias introduced by the missestimation of the strength of the
  H$\alpha$ absorption feature or equivalently of the continuum
  level. Black dots show show the fraction of the observed flux that
  we would observe if the stellar absorption was not considered at
  all. Red dots show the same fluxes corrected using a constant
  absorption EW=-2.4\AA. Dark blue and green dots correspond to
  understimations and overestimations of the continuum by a
  10\%. Light blue and orange dots correspond to
  understimations and overestimations of the continuum by a
  50\%.}
\label{fig-16}
\end{center}
\end{figure}

The observed line ratios imply that
assuming a constant NII/H$\alpha$ ratio throughout the galaxy would
introduce systematic overestimations of the H$\alpha$ flux of up to 40\%
in the faintest regions, as well as systematic underestimations of up
to 25\% for the brightest regions. The effect is a strong function
of H$\alpha$ flux, and its magnitude is of the order of the typical 
uncertainties quoted for narrow-band photometry of star-forming
regions in nearby galaxies. While in theory these systematic
missestimations should bias a measurement of the slope of the SFL
towards shallower values, the magnitude of the effect is ten times smaller
than the intrinsic scatter in the SFL and the introduced bias is 
negligible.

Now lets look at the effects introduced by errors in the continuum 
subtraction and estimation of the underlying H$\alpha$ stellar
absorption. When doing narrow-band imaging, the estimated value for the
H$\alpha$ absorption EW is coupled, and impossible to separate from the
estimated continuum level. So overestimations of the absorption EW can 
be thought as underestimations of the subtracted continuum and
viceversa. Black crosses in Figure \ref{fig-16} show the observed H$\alpha$
emission flux (before dust extinction correction) versus the
fractional difference between the H$\alpha$ emission and absorption
fluxes for all the regions unaffected by AGN contamination. The magnitude of the
H$\alpha$ absorption was measured in the best-fitted stellar continuum
spectrum of each region, constructed as described in Section 5.1. The
vertical axis in Figure \ref{fig-14} can be interpreted as the fraction of
the true flux we would observe if the underlying absorption was not
taken out from our measurement. Negative values correspond to regions
in which the absorption EW is higher than the emission EW. We measure
a fairly constant absorption EW, showing a median of -2.4\AA , and rms
scatter of 0.2\AA\ between different regions. This supports the
approximation of a constant H$\alpha$ absorption EW on which narrow-band
corrections are based. Not taking into account the absorption feature 
can translate into gross
underestimations of the emission line fluxes. For the brightest
regions the underestimation can be up to $\sim$50\%, and for the
faintest regions we could completely miss the presence of nebular
emission, and observe pure absorption.

The red crosses in Figure \ref{fig-16} show the emission minus
absorption fluxes corrected using a constant H$\alpha$ absorption EW
of -2.4\AA. It can be seen that, under the assumption of a constant
absorption EW, true fluxes can be recovered with typical uncertainties
of less than 20\% if the correct value of the median EW is
used. Green and blue crosses in Figure \ref{fig-16} correspond to the
values that would be obtained if the continuum had been
overestimated and underestimated by 10\% respectively, or
equivalently if the H$\alpha$ absorption EW  had been underestimated
by -0.2\AA\ and overestimated by +0.3\AA. The orange and light blue
crosses correspond to continuum misestimations of a 50\% (-0.8\AA,
+2.4\AA). These offsets are of the same order of magnitude as the
typical uncertainties in the continuum subtraction of narrow-band
images of nearby galaxies. It can be seen that a systematic
misestimations can be introduced to the measured H$\alpha$ fluxes,
especially in the fainter regions. Similarly to the [NII] correction
discussed above,
this effect is a strong function of H$\alpha$ flux and in this case
can induce a significant change in the slope of the SFL if the
estimated absorption (continuum level) is sufficiently off from the
true value. A 10\% error in the continuum level 
can introduce systematic misestimations of up to 30\%, which is
small compared to the intrinsic scatter in the SFL, but a 50\% error
in the estimation of the continuum can induce misestimations of the
measured fluxes that are of the order of the SFL intrinsic scatter,
and hence introduce a significant systematic bias to the SFL slope.

\begin{figure}[t]
\begin{center}
\epsscale{1.15}
\plotone{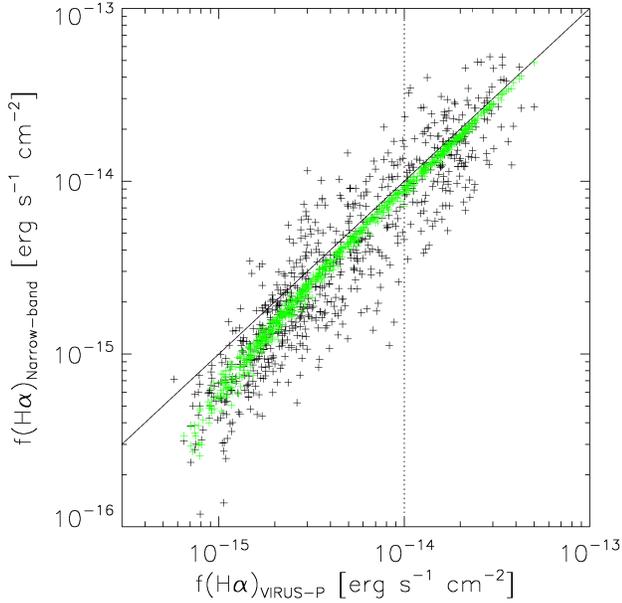}
\caption{VIRUS-P observed H$\alpha$ fluxes (before dust exticntion
  correction) versus H$\alpha$ fluxes measured in the continuum
  subtracted image from \cite{calzetti05} (balck crosses). Data-points
  to the right of the vertical dotted line were used to scale the
  narrow-band fluxes in order to account for flux calibration and
  apperture discrepancies. The green crosses show the H$\alpha$ fluxes
  that would have been measured by VIRUS-P if the continuum would have
been overestimated by a 30\% (see Figure \ref{fig-16}).}
\label{fig-17}
\end{center}
\end{figure}

We perform a comparison of our spectroscopically measured H$\alpha$
emission line fluxes to fluxes measured by performing photometry in
4.3$''$ diameter apertures at the positions of each of our fibers on
the continuum-subtracted and absorption line corrected narrow-band
image used by \cite{calzetti05} and \cite{kennicutt07}. We correct
the narrow-band fluxes for [NII] contamination using the correction
factors shown in Figure \ref{fig-13}, scaled by 0.97 to account for
the lower filter transmission at the [NII] lines. Figure \ref{fig-17}
shows the comparison. In order to account for differences in flux
calibration and photometry aperture effects, we scale the narrow-band 
fluxes by a factor of 1.25, given by the mean ratio between the VIRUS-P
and narrow-band fluxes for regions with $f(H\alpha)>10^{-14}$erg 
s$^{-1}$cm$^{-2}$ (to the right of the dotted line in Figure
\ref{fig-15}). At high H$\alpha$ emission fluxes the effects of errors in the
continuum subtraction are much smaller than for the fainter regions,
so we consider safe to scale the fluxes in order to match the bright
end of the distribution, also the magnitude of the scaling factor is
of the order of the combined uncertainties in flux calibration.

Narrow-band fluxes presented in Figure \ref{fig-17} should not be
affected by previously discussed systematics introduced by [NII]
corrections, since we used the spectroscopically measured ratios to
correct them. On the other hand, they clearly show a systematic
deviation, with narrow-band fluxes being lower than spectroscopic
fluxes as we go to fainter regions. This is consistent with an
overestimation of the continuum level by $\sim$30\%, or equivalently
and underestimation of the H$\alpha$ absorption EW by -0.6\AA, which
is well within the uncertainties involved in the continuum subtraction
of the narrow-band image (Calzetti private comunication). It is
important to notice that in \cite{kennicutt07}, the spatially resolved
SFL was built by doing photometry on H$\alpha$ bright
star-forming knots (brighter than $3\times10^{-15}$ erg
s$^{-1}$cm$^{-2}$), which are less affected by errors in the continuum
subtraction than for example the inter-arm regions. Hence we do not
expect this effect to significantly affect the slope of the SFL
that they measure.

The above comparison stresses a very important point. Although very
deep narrow-band imaging can be obtained using present day imagers,
low surface-brightness photometry of nebular emission in these images is limited by
uncertainties in the continuum subtraction and estimation of photospheric
absorption. In this respect, integral field spectroscopy provides us
with a less biased way of measuring faint nebular emission in nearby
galaxies.

\section{Comparison with Previous Measurements and Theoretical Predictions}

In this section we compare our results to the recent measurements on
the spatially resolved SFL in NGC5194 by \cite{kennicutt07} and
\cite{bigiel08}, and to the predictions of the theoretical model of
the SFL proposed by \cite{krumholz09b}.

We find an almost complete lack of correlation between the atomic gas
surface density and the SFR surface density (Figure
\ref{fig-10}). This is in good agreement with the observation of both
\cite{kennicutt07} and \cite{bigiel08}, and confirms the fact that the
SFR is correlated with the molecular gas density, and it is this
correlation which drives the power-law part of the total gas SFL.
At low gas surface densities ($<20$M$_{\odot}$pc$^{-2}$) the ISM of spiral
galaxies stops being mostly molecular, and hence the shape of the 
total gas SFL is driven by a combination of the molecular gas SFL and 
the ratio of molecular to atomic hydrogen.

As discussed in \S 1, \cite{kennicutt07} finds a super-linear slope
of $1.37$ for the molecular SFL in NGC5194, while \cite{bigiel08}
measures a slightly sub-linear slope of $0.84$. The first of these
measurements is consistent with models in which the SFR is inversely
proportional to the gas free-fall time in GMCs and the molecular gas
surface density is proportional to the total gas density ($N=1.5$,
\cite{kennicutt98b}), while the second is more consistent with models
in which the SFR shows a linear correlation with the molecular gas
density, product of star-formation taking place at a constant
efficiency in GMCs. Hence, establishing the slope of the SFL is
important in order to distinguish between different physical phenomena that
give rise to it.

\begin{figure}[t]
\begin{center}
\epsscale{1.2}
\plotone{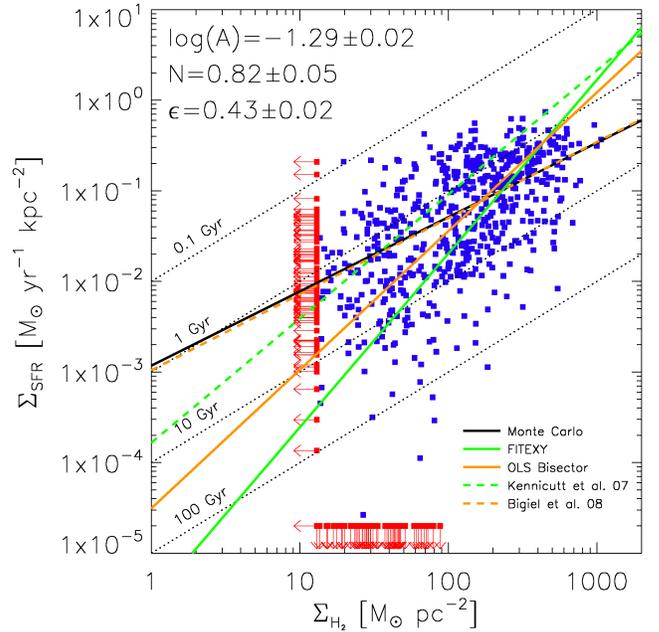}
\caption{Molecular gas SFL as measured by VIRUS-P. Symbols are the
same as in Figure \ref{fig-12}. The black solid line shows our best
fitted power-law obtained using the Monte Carlo method described in \S
9.1. Previous measurements by \cite{kennicutt07} and \cite{bigiel08}
are shown as the green and orange dahsed lines respectively. Also
shown are fits to our data (rejecting upper limits) using the FITEXY (solid green line) and OLS
bisector (solid orange line) methods.}
\label{fig-18}
\end{center}
\end{figure}

Figure \ref{fig-18} shows the molecular SFL measured as described in
\S 9, together with the best-fitted SFL as measured by \cite{kennicutt07}
and \cite{bigiel08}. The results from the latter are adjusted to
account for differences in the IMF assumed for calculating
$\Sigma_{SFR}$, and the different CO-H$_2$ conversion factor used in the
calculation of $\Sigma_{H_2}$. 

Our best-fitted molecular SFL shows a
considerably shallower slope than the one measured by
\cite{kennicutt07}. We consider the source of the disagreement to be a
combination of two factors. First, as shown in \S 10, the narrow-band
H$\alpha$ fluxes used by \cite{calzetti05} and \cite{kennicutt07}
might be underestimated at the faint end of the flux distribution due
to small systematic errors in continuum subtraction, although the
effect is small (of the order of the intrinsic scatter in the SFL),
and cannot account for the bulk of the difference observed in the SFL
slope. The second factor, which we consider to be the main cause
behind the disagreement, is the difference in the
fitting methods used to adjust a power-law to the
data. As mentioned in \S 9, \cite{kennicutt07} used a FITEXY algorithm
to perform a linear regression to the data in logarithmic space,
rejecting upper limits in $\Sigma_{H2}$ from the fit, and not fitting
for the intrinsic scatter in the SFL. The solid green
line in Figure \ref{fig-17} shows the result of applying the same
procedure to our data. The FITEXY method significantly overpredicts 
the slope of the SFL ($N=1.9$), in large part due to the exclusion of
the $\Sigma_{H2}$ upper limits. These data-points, having large
error bars in $\Sigma_{gas}$ and clear detections in $\Sigma_{SFR}$,
have a significant statistical weight in the Monte Carlo
fit because of their large number. Another factor promoting the
fitting of shallower slopes by our Monte Carlo method, is the fact that
we included the intrinsic dispersion in the SFL as a scatter in
$\Sigma_{SFR}$, hence the fit will tend to equalize the number of
data-points above and below the power-law at any given
$\Sigma_{gas}$. This is a consequence of the expectation for a causal
relation between $\Sigma_{gas}$ and $\Sigma_{SFR}$, with the SFR
beeing a function of the gas density, and not
viceversa. 

\cite{kennicutt07} provide a table of their measured values for
$\Sigma_{SFR}$ and $\Sigma_{HI+H_2}$ and their uncertainties, from
which they recover a slope of $N=1.56$ for the total gas SFL. We apply
our Monte Carlo fitting method to their data, and find best-fitted
values of $A=10^{-1.23\pm0.03}$ for the amplitude,
$\epsilon=0.40\pm0.03$ for the intrinsic scatter, and a slope
$N=1.03\pm0.08$. This shallower slope is a lot closer to our mesured
value of $N=0.85$, and the rest of the difference can be easily
explained by the underestimation of the narrow-band H$\alpha$ fluxes
presented in Figure \ref{fig-17} and differences in the DIG
correction. The two independent datasets show excellent agreement in 
the value of intrinsic scatter. The small difference of 0.08 dex in the
amplitude can be attributed to the fact that
\cite{kennicutt07} targeted active star-forming regions in their
study,  and hence their measurement of the SFL is most likely biased 
towards higher star-formation efficiencies than the one presented here.

On the other hand, we measure a molecular SFL which shows an excellent
agreement with \cite{bigiel08} both in slope and normalization. The
agreement is better than expected, given the differences in the
methods used to measure $\Sigma_{SFR}$ and fitting the SFL. Their SFR
measurements are not based on extinction corrected hydrogen
recombination lines as in Kennicut et al. and this work, but rather on
a linear combination of space-based GALEX far-UV and $Spitzer$ MIPS
24$\mu$m fluxes. Also, they do not correct their data in order to
account for any contribution from the DIG. The fitting method used by 
\cite{bigiel08} is an OLS Bisector, and they also reject non
detections in CO from the fit. The orange solid line shows the result
of applying this fitting method to our data. Just as in the case of
the FITEXY algorithm, the OLS Bisector yields a significantly higher 
slope ($N=1.5$) than the Monte Carlo fit. The reasons for this are the
same as for the FITEXY algorithm, that is, the inclusion of the upper
limits in $\Sigma_{H_2}$, and the introduction of the intrinsic
scatter in $\Sigma_{SFR}$ in our method. One possible explanation for
the agreement could be the interplay between the lack of DIG
correction and the difference in fitting methods. The first will tend
to drive the slope to shallower values, while the second will steepen
it. The combination of these two effects working in opposite
directions might be behind the agreement between \cite{bigiel08} and
this work. 

Although the comparison is hard due to the systematics involved in the 
different methods, the bottom line is that we measure a slope that is
consistent with the scenario proposed by \cite{bigiel08} and
\cite{leroy08}, in which star-formation takes place at a nearly constant
efficiency in GMCs over a large range of environments present in
galaxies. This is also in agreement with recent the findings of
\citep{bolatto08}, who find that extragalactic GMCs in the Local
Group, detected on the basis of their CO emission, exhibit remarkably
uniform properties, with a typical mass surface density of roughly 85
M$_\odot$pc$^{-2}$.

\begin{figure}[t]
\begin{center}
\epsscale{1.15}
\plotone{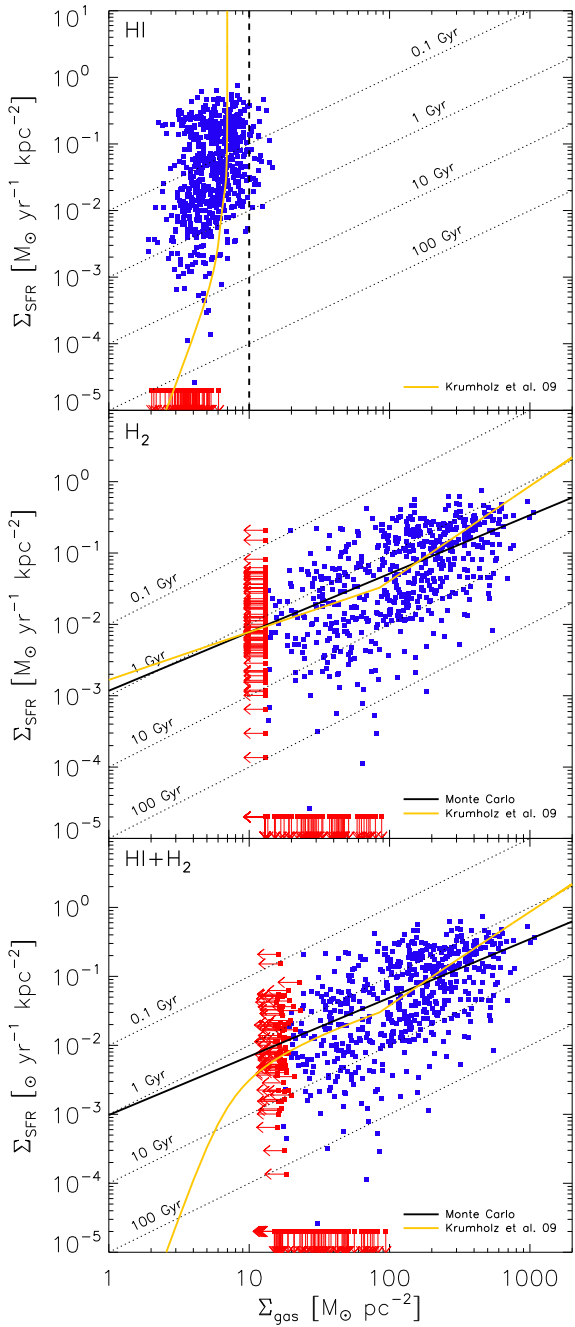}
\caption{Comparison of the observed SFL for atomic gas (top),
  molecular gas (center), and total gas (bottom) and the theoretical
  model proposed by \cite{krumholz09b}. Symbols are the same as in
  Figures \ref{fig-11}, \ref{fig-12}, and \ref{fig-12}. The solid
  orange line show the Krumholz et al model for $Z'=1.0/1.5$ and $c=4$.}
\label{fig-19}
\end{center}
\end{figure}

Based on these concepts of uniformity of GMC properties, and good
correlation between the SFR and the molecular gas density,
\cite{krumholz09b} proposed a simple theoretical model to explain the
observed total gas SFL. In their model, star formation takes place
only in molecular gas, and the total gas SFL is
determined by three factors. First, the fraction of the gas in
molecular form is set by the balance between the formation of
H$_2$ in the surface of dust grains, and the dissociation of molecules
by the far-UV continuum in the Lyman-Werner bands \citep{krumholz08,
 krumholz09a}. This drives the
shape of the total gas SFL in the low density regime where the ISM is not fully
molecular. Second, the star-formation efficiency inside GMCs is low,
and it is set by turbulence driven feedback processes \citep{krumholz05}. These are
responsible for the power-law behavior of the molecular SFL.
Third, GMCs are decoupled from the surrounding ISM when
their internal pressure is higher than external
pressure. In this regime their structure is determined by internal
feedback processes, and they show very uniform properties including an almost
constant surface density of 85
M$_\odot$pc$^{-2}$ \citep{bolatto08}. When the galactic ISM pressure
becomes higher than this value, the GMC surface density must increase
accordingly in order to maintain pressure balance with the external
ISM. This gives rise to a steepening of the slope of the molecular SFL at
$\Sigma_{H_2} \ge 85 \rm{M_{\odot}pc^{-2}}$. In summary, the total gas
SFL in the model shows a different behavior in the low,
intermediate, and high density regimes. At low densities its
behavior is driven by the transition from an atomic to a molecular
ISM. Beyond the point at which the ISM becomes almost fully molecular
the total gas SFL  follows closely the molecular SFL, which shows a
steeper slope in the high density regime driven by the pressure
balance between the galactic ISM and GMCs.

Figure \ref{fig-18} shows a comparison of our data and the
\cite{krumholz09b} model. We have assumed $Z'=Z/Z_{MW}=1.0/1.5$,
consistently with the DIG correction applied in \S 8, and a clumpiness
factor $c=4$ to account for the effect that the averaging of
$\Sigma_{gas}$ introduces in the molecular fraction in the model. We
observe an excellent agreement for both the atomic and molecular gas,
as well as for the total gas SFL. The gas density range
sampled by our observations, and the scatter in SFL does not allow us
to discern between the model and the simple power law fitted using the
Monte Carlo method, stressing the need to extend our observations
towards more extreme density environments. 

\section{Summary and Conclusions}

We have performed the first measurement of the spatially resolved SFL
in nearby galaxies using integral field spectroscopy. The wide field
VIRUS-P spectroscopic map of the central $4.1\times 4.1$
kpc$^2$ of NGC 5194, together with the HI 21cm map from THINGS, and
the CO J=1-0 from BIMA SONG were used to measure $\Sigma_{SFR}$,
$\Sigma_{HI}$, and $\Sigma_{H_2}$ for 718 regions $\sim$170 pc in
diameter throughout the disk of the galaxy. 

In this paper we have
presented our method for calculating $\Sigma_{SFR}$ from the
spectroscopically measured H$\alpha$ emission line fluxes. We have shown
that the observed H$\alpha$/H$\beta$ ratio is a good estimator of the
nebular dust extinction, at least at the levels of obscuration present
in face-on normal spiral galaxies like NGC 5194. 

We have also
presented a new method
for estimating the contribution of the DIG to the H$\alpha$ emission
line flux, which is based on the observed low-ionization line ratio
[SII]/H$\alpha$, and the large differences seen in this line ratio
between H II regions and pointings towards the DIG in the
Milky Way. The use of line ratios to correct both for dust extinction and
the DIG contribution is possible only because of the use of integral
field spectroscopy spanning a large wavelength range, which includes
all these important emission lines.

One of the main goals of this work is to make use of these clean
spectroscopic emission line measurements to study the systematics involved in
narrow-band estimations of the H$\alpha$ emission line flux of nearby
galaxies. We showed that proper estimation of the continuum and of the
underlying stellar absorption features is crucial in order to get an
unbiased estimate of the H$\alpha$ flux. Errors of the order of
30\% in the estimation of these quantities can introduce
systematic misestimations of the H$\alpha$ emission line flux by up 
to a factor of 3 in the low surface brightness regime.

We also tested the assumption of a constant [NII]/H$\alpha$ ratio
throughout the galaxy, usually used to remove the [NII] doublet
contamination from the narrow-band measured fluxes. We found that the
[NII]/H$\alpha$ ratio varies significantly throughout the galaxy, and
shows a clear correlation with the H$\alpha$ flux. The sense of the 
correlation implies a higher [NII]/H$\alpha$ ratio in regions that 
are fainter in H$\alpha$ (typically the inter-arm regions of the
galaxy), and is consistent with the DIG dominating the nebular
spectrum in these zones. Assuming a constant [NII]/H$\alpha$ would
introduce overestimations of the H$\alpha$ flux of $\sim$40\% in the
inter-arm regions, and underestimations of $\sim$25\% for the brightest
star-forming regions in the spiral arms.

Integral field spectroscopy proves to be an extremely powerful tool
for mapping the SFR throughout the disks of nearby galaxies,
especially with the advent of large field of view IFUs like VIRUS-P.
Spatially resolved spectral maps, besides allowing us to measure
emission line fluxes in a much more unbiased way than narrow-band
imaging, also provides extensive information about the physical
conditions throughout the disks of nearby spiral galaxies. The spectra
allows the measurement of metallicities, stellar and gas kinematics,
stellar populations, and star formation histories across galaxies. In 
a future study we will investigate the role that all these other
quantities that can be extracted from our data play at setting the SFR.

We found that the SFR surface density shows a lack of correlation
with the atomic gas surface density, and a clear correlation with the
molecular gas surface density. Hence, the total gas SFL is fully
driven by the molecular gas SFL in the density regimes sampled by our 
observations. The atomic gas surface density is observed to saturate
at a value of $\sim$10 M$_{\odot}$pc$^{-2}$, at which a phase
transition between atomic and molecular gas is thought to occur in
the ISM.

A Monte Carlo method for fitting the SFL which is not affected by the
systematics involved in performing linear correlations of incomplete
data in logarithmic space was presented. Our method fits the intrinsic
scatter in the SFL as a free parameter. Applying this method to our
data yields slightly sub-linear slopes $N$ of 0.82 and 0.85 for the 
molecular and total gas SFLs respectively. 

Comparison with previous
measurements of the spatially resolved SFL are somewhat challenging
because of the different recipes used to estimate $\Sigma_{SFR}$, and
the different fitting procedures used to derive the SFL
parameters. The slopes we measured are in disagreement with the
results of \cite{kennicutt07}, who measured a strongly super-linear
slope for both the molecular component and the total gas. On the other
hand, our results are in very good agreement with the slope measured
for the molecular gas SFL in NGC 5194 by \cite{bigiel08}. 
Our results are consistent with the scenario recently proposed by
\cite{bigiel08} and \cite{leroy08} of a nearly constant SFE in GMCs,
which is almost independent of the molecular gas surface density. The main
argument to support this scenario is the observation of a close to
linear correlation between the $\Sigma_{SFR}$ and $\Sigma_{gas}$ in
the density ranges present in the ISM of nearby normal spiral
galaxies. 

On the other hand our results also show a very good agreement with the
more complex scenario recently proposed by \cite{krumholz09b}, in
which the surface density of molecular gas grows with the
molecular to atomic fraction at low densities ($\Sigma_{HI+H_2}\lesssim$10
M$_{\odot}$pc$^{-2}$), becomes constant at intermediate densities (10
M$_{\odot}$pc$^{-2}\lesssim \Sigma_{HI+H_2}\lesssim$100
M$_{\odot}$pc$^{-2}$), and increases linearly with the total gas
density in the high density regime ($\Sigma_{HI+H_2}\gtrsim$100
M$_{\odot}$pc$^{-2}$). This, combined with an slightly sub-linear
efficiency as a function of molecular gas surface density given by the
balance between gravitational potential energy and turbulent kinetic
energy originated by internal feedback, gives rise to the observed
SFL. In their model, the total gas SFL has a super-linear slope $N=1.33$ 
in the high density regime, gets shallower at intermediate densities
showing a slope of $N=0.67$, and steepens again at lower densities as
the molecular to atomic gas fraction rapidly decreases. Our
observations sample the transition between the intermediate and high
density regimes in the model. The intrinsic scatter in the SFL,
together with our limited density dynamic range does not allow us
to observe the predicted kink in the SFL directly, but our measured
slope of 0.85 is very close to what we expect to measure in a region
where we sample both the sub-linear and super-linear parts of the SFL
predicted by Krumholz et al. model. A proper detection of the kink in 
the SFL predicted by \cite{krumholz09b} will require extending the 
dynamic range to higher gas surface densities. 

A major success of the \cite{krumholz09b} model is the excellent
agreement it shows with the observation with respect to the SFE, or
equivalently to the gas depletion timescales. We observe very long
depletion timescales of $\tau \approx$2 Gyr, in good agreement with
previous observations. This time is $\sim$100 longer than the typical
GMC free-fall time. The good agreement between our observations and the
Krumholz et al. model implies that this very low efficiency can be
easily explained by models in which star-formation is self regulated
through turbulence induced by internal mechanical feedback in GMCs.

An important result of this study is the large intrinsic scatter
of 0.43 dex observed in both the molecular and total gas SFLs. This
translates into a factor of $\sim$3 scatter  
in the SFR for regions having the same molecular gas availability, and
it may indicate the existence of further parameters that are
important in setting the SFR. It is worth mentioning that part of the 
intrinsic scatter in the SFL must come from the scatter in the SFR-L(H$\alpha$)
calibration. \cite{charlot01} show that SFRs derived from H$\alpha$
alone present a large scatter when compared to SFRs derived from full
spectral fitting of the stellar populations and nebular emission of
a sample of 92 nearby star-forming galaxies. Recently, the detection 
of widespread UV emission beyond the H$\alpha$ brightness profile
cutoff in the outer disks of many nearby galaxies \citep{gildepaz05,
  thilker05, boissier07} , has raised questions 
about the proportionality between the H$\alpha$ emission and the SFR
in the low star-formation regime. Incomplete sampling of the IMF in
low-mass embedded clusters has been proposed to explain the
discrepancy between H$\alpha$ and UV surface brightness profiles
\citep[e.g.][]{pflamm09}. Under this scenario the H$\alpha$ emission fails to
tracing star-formation in low mass clusters where statistical
fluctuations can translate into a lack of massive ionizing stars, and
the SFR-L(H$\alpha$) becomes non-linear in the low star-formation
regime \citep{pflamm07}, which might enhance the downward scatter in
our SFL measurements. This issue is beyond the scope of the current paper,
but we intend to investigate the implications of applying non-linear
SFR-L(H$\alpha$) to our data in future works.

In this paper we have established the method for studying the
spatially resolved SFL using wide integral field
spectroscopy, and have set new constrains on important quantities like the
slope, normalization, and intrinsic scatter of the SFL. As mentioned in \S 1, this data forms part
of an undergoing large scale IFU survey of nearby galaxies. VENGA will
map the disks of $\sim$20 nearby spiral galaxies to radius much larger
than those sampled by the data presented here. In the future, we will
extend this type of study to a larger set of galaxies spanning a
range in Hubble types, metallicities, and star-formation
activities. This will help us to sample a larger dynamical range in
gas surface densities. The later requires
the observation of much denser environments, like the ones present in
starburst galaxies, to extend the observed SFL to higher
densities. Deeper CO observations that map the molecular gas out
to large radii will be necessary to extend the sampled range to lower
densities. This is of great importance, since a proper
characterization of the shape of the total gas SFL is necessary in
order to distinguish between different star-formation models.\\

We thank Phillip McQueen and Gary Hill for designing and constructing
VIRUS-P, and for their advice on the use of the instrument. We also
acknowledge David Doss and the staff at McDonald Observatory for their
invaluable help during the observations. Tables of the WHAM DIG line
ratios were kindly provided by George J. Madsen. We thank Daniela
Calzetti for some very useful discussions, and for providing the
narrow-band Pa$\alpha$ and H$\alpha$ images used in her work. This
study has been possible thanks to the financial support of the Sigma 
Xi, The Scientific Research Society. The construction of VIRUS-P was 
possible thanks to the generous support of the Cynthia \& George
Mitchell Foundation. NJE and AH were supported in part by NSF Grant
AST-0607193. Finally we thank the referee for the helpful comments
which helped to improve the quality of this work.

\clearpage

\LongTables
\begin{landscape}

\begin{deluxetable}{lcccccccccc}
\tabletypesize{\scriptsize}
\tablecaption{Nebular Emission Line Fluxes, Gas Surface Densities, and
SFR Surface Densities}
\tablewidth{0pt}
\tablehead{
\colhead{ID} & \colhead{Equatorial Coordinates} & 
\colhead{H$\beta$} & \colhead{[NII]$\lambda$6548} &
\colhead{H$\alpha$} & \colhead{[NII]$\lambda$6584} &
\colhead{[SII]$\lambda$6717} & \colhead{[SII]$\lambda$6731} &
\colhead{$\Sigma_{H2}$} & \colhead{$\Sigma_{HI}$} &
\colhead{$\Sigma_{SFR}$}
}

\startdata
& $\;\;\;\;\alpha\;\;\;\;\;\;\;\;\;\;\;\;\;\;\;\;\delta$ & 
& & 
$\;\;\;\;\;\;10^{-16}$ erg s$^{-1}$cm$^{-2}$ & & 
 & &
$M_{\odot}pc^{-2}$ & $M_{\odot}pc^{-2}$ & $M_{\odot}yr^{-1}kpc^{-2}$\\

\tableline
\\

1 & 13:29:48.11 +47:12:35.6 & 12.57$\pm$0.11 & 6.45$\pm$0.42 & 69.73$\pm$0.59 & 17.63$\pm$0.43 & 6.89$\pm$0.23 & 4.95$\pm$0.23 	 & 10.71$\pm$13.00 & 9.75$\pm$0.59 & 0.08181$\pm$0.01652 \\
2 & 13:29:48.83 +47:12:35.5 & 4.22$\pm$0.41 & 4.94$\pm$0.46 & 29.44$\pm$0.54 & 13.90$\pm$0.48 & 4.77$\pm$0.43 & 3.12$\pm$0.43 	 & 127.44$\pm$13.00 & 12.25$\pm$0.59 & 0.05385$\pm$0.01664 \\
3 & 13:29:49.55 +47:12:35.3 & 6.05$\pm$0.25 & 5.89$\pm$0.43 & 36.35$\pm$0.52 & 17.82$\pm$0.47 & 6.02$\pm$0.10 & 4.63$\pm$0.10 	 & 399.25$\pm$13.00 & 8.73$\pm$0.59 & 0.04681$\pm$0.01063 \\
4 & 13:29:50.28 +47:12:35.2 & 4.78$\pm$0.25 & 4.42$\pm$0.53 & 27.71$\pm$0.59 & 13.85$\pm$0.55 & 4.11$\pm$0.10 & 3.02$\pm$0.10 	 & 174.20$\pm$13.00 & 10.05$\pm$0.59 & 0.02975$\pm$0.00752 \\
5 & 13:29:51.00 +47:12:35.0 & 2.94$\pm$0.01 & 4.88$\pm$0.55 & 16.37$\pm$0.54 & 11.66$\pm$0.54 & 3.19$\pm$0.03 & 2.44$\pm$0.03 	 & 127.30$\pm$13.00 & 10.53$\pm$0.59 & 0.01123$\pm$0.00274 \\
6 & 13:29:51.72 +47:12:34.8 & 4.75$\pm$0.04 & 5.77$\pm$0.29 & 24.74$\pm$0.31 & 16.43$\pm$0.30 & 4.64$\pm$0.10 & 2.99$\pm$0.10 	 & 131.55$\pm$13.00 & 7.08$\pm$0.59 & 0.01791$\pm$0.00369 \\
\end{deluxetable}
\clearpage
\end{landscape}

\end{document}